\documentclass[11pt]{article}
\usepackage[margin=1in]{geometry}
\usepackage[T1]{fontenc}
\usepackage{graphicx}      
\usepackage{booktabs}      
\usepackage{multirow}      
\usepackage{array}         
\usepackage{tabularx}      
\usepackage{float}         
\usepackage{placeins}      
\usepackage{amsmath}       
\usepackage{amssymb}       
\usepackage{adjustbox}     
\usepackage{xcolor}
\usepackage{algorithm}
\usepackage{algpseudocode} 
\usepackage[numbers,sort&compress]{natbib}
\usepackage{hyperref}      
\hypersetup{
  colorlinks=true,
  linkcolor=blue,
  citecolor=blue,
  urlcolor=blue
}

\algrenewcommand\algorithmicrequire{\textbf{Input:}}
\algrenewcommand\algorithmicensure{\textbf{Output:}}

\renewcommand{\algorithmicrequire}{\textbf{Input:}}
\renewcommand{\algorithmicensure}{\textbf{Output:}}
\begin{document}
\title{NeuRoute: Logit-Guided Neural Routing for Billion-Scale Vector Search with Sub-Hour Index Construction}

\author{
Xingqiao Wang \quad Zi Wang \quad Xiaowei Xu\thanks{Corresponding author. Email: \texttt{xwxu@ualr.edu}}\\
\small \texttt{xingqiao078@gmail.com} \quad \texttt{zwang@ualr.edu} \quad \texttt{xwxu@ualr.edu}\\
\small University of Arkansas at Little Rock
}
\date{}

\maketitle

\begin{abstract}
Building approximate nearest neighbor (ANN) indexes at billion
scale is often dominated by expensive global clustering or graph
construction, making time-to-index a first-order systems concern.
We present \emph{NeuRoute}, a learned hashing index that turns short binary codes into an effective routing primitive for large-scale vector search. \emph{NeuRoute} trains a lightweight neural network encoder with a selective similarity-preserving objective to produce well-balanced binary
addresses. During construction, \emph{NeuRoute} organizes vectors into buckets by their codes and performs \emph{bucket-local} clustering in the encoder’s low-dimensional space to form centroids. At query time, \emph{NeuRoute} exploits the encoder logits as an uncertainty signal: it uses deviation-to-threshold scores to prioritize uncertain-bit perturbations for query-adaptive multi-bucket probing, scores \emph{bucket-local} centroids by their distances to the query to form a compact
candidate cluster set, and applies centroid-stage gating with heap-quality-driven early stopping to prune low-value clusters before exact refinement.
On billion-scale benchmarks, \emph{NeuRoute} achieves strong accuracy--throughput trade-offs with fast index construction: on BigANN-1B it reaches 90.3\% Recall@10 at 2{,}414 QPS and is 1.7$\times$ faster than OPQ+IVF-PQ (refine) at comparable accuracy, while completing end-to-end training+construction in under an hour on both BigANN-1B and Deep1B-1B.
These results show that logit-guided neural routing can make hashing competitive as a lightweight ANN indexing framework at billion scale.
Source code and artifacts are available at \url{https://github.com/XingqiaoWang/NeuRoute}.
\end{abstract}

\section{Introduction}
\label{sec:introduction}

\begin{figure}[t]
  \centering
  \includegraphics[width=\linewidth]{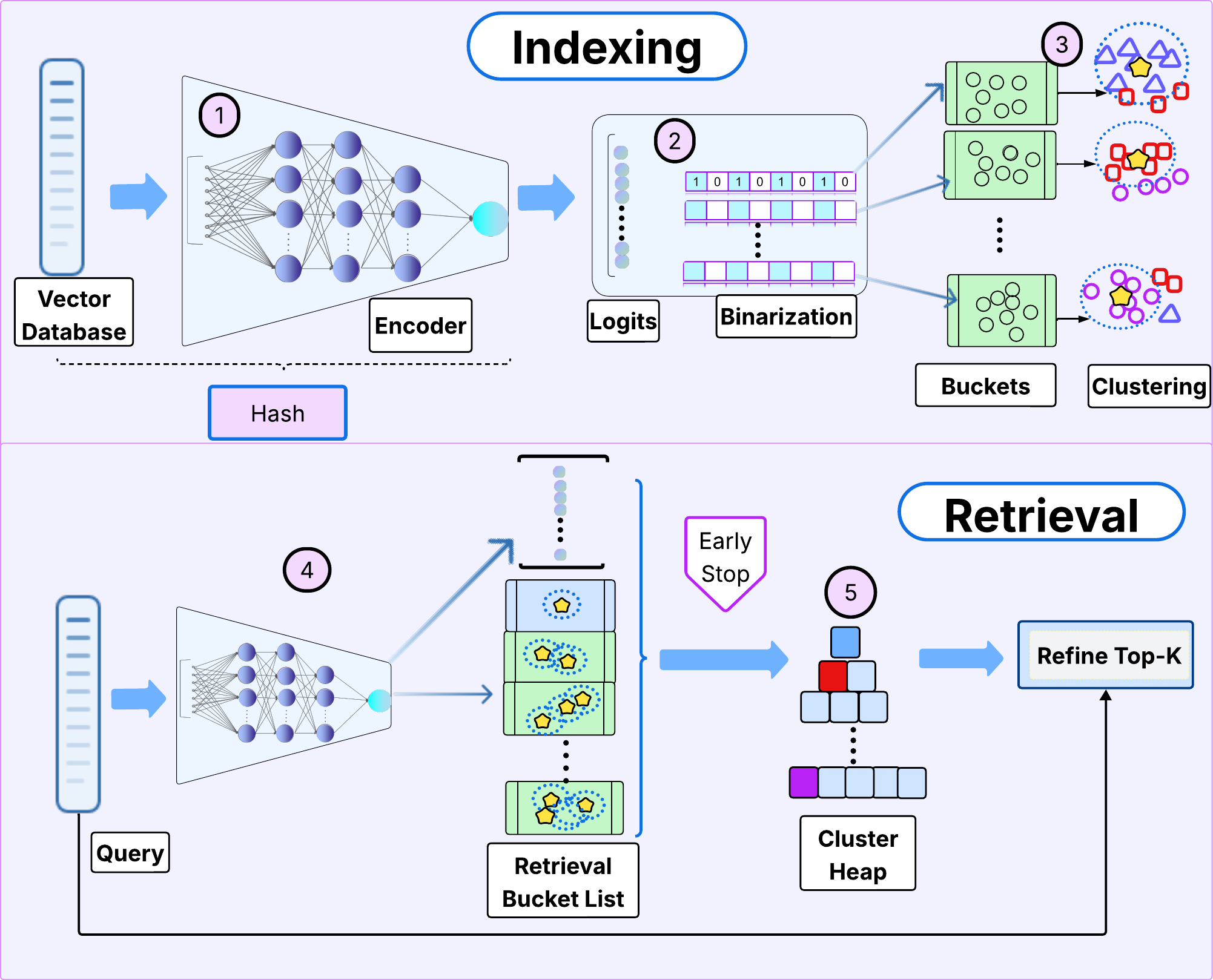}
  \caption{Overview of the \textit{NeuRoute} pipeline (Steps~1--5). Steps~1--3: hash learning and index construction; Steps~4--5: logit-guided candidate generation with calibrated gating and early stopping before exact scoring.}
  \label{fig:spoverview}
\end{figure}

Billion-scale similarity search is a cornerstone of modern data management and machine learning systems, supporting applications such as image retrieval, recommendation, deduplication, and large-scale embedding search~\cite{ref15,10.1145/3448016.3457550,simhadri2022bigann}.
To meet strict latency and memory constraints, practitioners rely on approximate nearest neighbor (ANN) indexes, with graph-based methods and disk-resident hybrids offering strong recall--latency trade-offs on large datasets~\cite{33,NEURIPS2019_09853c7f,wang2021comprehensivesurveyexperimentalcomparison,NEURIPS2021_299dc35e}.
However, at billion scale these approaches often incur substantial system costs---large index footprints, expensive construction and tuning, and heavy preprocessing such as global clustering or graph building---which become first-order concerns under single-node resource budgets~\cite{wang2021comprehensivesurveyexperimentalcomparison,10.14778/3594512.3594527,NEURIPS2019_09853c7f}.

Hashing provides a complementary design point.
Binary codes are compact, inexpensive to store, and fast to compute, enabling lightweight partitioning of massive databases into buckets (lists) that can be scanned efficiently~\cite{NIPS2008_d58072be,inproceedings2011,Liong_2015_CVPR}.
Yet, despite extensive progress in learning better codes, hashing has struggled to become a dependable routing mechanism at billion scale under tight scan budgets~\cite{10.14778/3594512.3594527,26}.
A core limitation is that many pipelines expand search using discrete Hamming-radius probing (e.g., radius $0/1/2/\ldots$) with weak prioritization, which provides limited control over where computation is spent~\cite{NIPS2008_d58072be,inproceedings2011,NIPS2012_59b90e10}.
Moreover, after binarization many methods discard pre-binarization activations (logits) that encode bit uncertainty---precisely the signal needed to prioritize probes and to terminate early when additional scanning is unlikely to improve results.

In this work, we revisit hashing from a systems perspective and ask:
\textbf{Can short binary codes become a dependable neural router for billion-scale retrieval under a strict scanning budget?}
We propose \textbf{NeuRoute}, an unsupervised neural hashing framework designed explicitly for routing efficiency and budget-controlled candidate generation.
Figure~\ref{fig:spoverview} sketches the end-to-end pipeline: Steps~1--2 learn a lightweight encoder that produces short binary addresses (\S\ref{subsec:hash-function}); Step~3 builds a bucketized index and injects additional selectivity via \emph{bucket-local} clustering in the learned low-dimensional space (\S\ref{subsec:index-build}); and at query time, Steps~4--5 perform logit-guided bucket enumeration and centroid-stage selection with calibrated gating and heap-quality-driven early stopping (\S\ref{subsec:retrieval}, Algs.~\ref{alg:bucket_enum} and~\ref{alg:centroid_enum_gate_simpleB}).
Concretely, NeuRoute uses deviation-to-threshold scores derived from encoder logits to prioritize uncertain-bit perturbations for query-adaptive multi-bucket probing, then ranks bucket-local centroids by distance to form a compact candidate cluster set.
A calibrated centroid-stage gate together with heap-quality-driven early stopping prunes low-value clusters, after which exact scoring in the original embedding space is used only for final top-$k$ refinement.

NeuRoute is built around a simple systems observation: while short binary codes are attractive for compact routing and fast bucketization, hashing-only retrieval remains insufficient at billion scale under strict scan budgets.
To make short codes useful for ANN without heavy offline pair mining or graph dependencies, NeuRoute trains a lightweight encoder with a selective similarity-preservation objective computed on-the-fly within mini-batches (\S\ref{subsec:hash-function}), avoiding expensive offline neighbor-graph construction and integrating naturally into frequent re-indexing workflows.
However, even with prioritized bucket enumeration, routing with short codes alone can still yield excessive refinement workload at practical recall levels.
This motivates NeuRoute's second-stage structure: lightweight bucket-local clustering in the learned low-dimensional space (\S\ref{subsec:index-build}), which adds centroid organization within each bucket to support aggressive candidate reduction before exact refinement.

Importantly, this structure is both feasible and impactful at billion scale: bucket-local clustering completes in minutes as part of Add/Build (e.g., 380\,s on BigANN-1B and 614\,s on Deep1B-1B) while improving end-to-end throughput by up to $4.57\times$ and reducing refinement candidates by up to $4.68\times$ at matched Recall@10 (Table~\ref{tab:ablation_bucket_clustering}).
To further keep query-time work budget-controlled, NeuRoute applies a calibrated centroid-stage gate together with a heap-quality--driven early-stop criterion (\S\ref{subsec:retrieval}), yielding an additional $1.5\times$--$1.7\times$ QPS gain with negligible recall loss (Table~\ref{tab:ablation_early_stop}).
Finally, we implement logit-guided bucket enumeration as an efficient engineering primitive that operationalizes uncertainty-aware probing, allowing the system to concentrate effort on high-yield buckets and stop early when additional centroid evaluations no longer improve the retained set.
Together, these components turn learned hashing from a strong but insufficient primitive into a fast-build, budget-controlled billion-scale retrieval system, enabling sub-hour end-to-end indexing time (0.82\,h/0.93\,h on BigANN-1B/Deep1B-1B) on a single node (\S\ref{sec:build-time}, Table~\ref{tab:NeuRoute_time_breakdown}).

NeuRoute is most closely related to learned hashing and discrete representations, but it is designed for billion-scale routing: a lightweight training phase produces short binary codes for Hamming-space navigation and efficient candidate generation.
Compared to graph- and disk-backed methods, NeuRoute reduces reliance on complex graph maintenance and storage/I/O-specific tuning.
Compared to IVF--PQ pipelines, it avoids global coarse clustering over the full dataset and heavyweight quantizer training, relying instead on bucketized indexing with lightweight bucket-local clustering in low dimension.
Overall, NeuRoute targets scalable retrieval with a simple index structure, fast time-to-index, and budget-controlled query-time compute through routing, centroid-stage filtering, and exact refinement.

\paragraph{Contributions.}
We make the following contributions:
\begin{itemize}
  \item \textbf{Logit-guided neural routing with short codes.}
  We propose NeuRoute, an unsupervised MLP-based hashing framework that produces short, well-balanced binary addresses tailored for large-scale routing.

  \item \textbf{ANN-aligned, lightweight training objective.}
  We introduce an unsupervised objective that aligns the encoder with ANN neighborhoods while remaining lightweight to train.

  \item \textbf{Bucket-local clustering for centroid-stage filtering.}
  We introduce bucket-local clustering in the learned low-dimensional space to add lightweight centroid structure within buckets.

  \item \textbf{Billion-scale evaluation and fast time-to-index.}
  We evaluate two 1B-scale benchmarks (BigANN-1B and Deep1B-1B) and a high-dimensional setting. NeuRoute provides strong Recall@10--throughput trade-offs and sub-hour end-to-end rebuilds (training $+$ Add/Build) on a single node.
\end{itemize}

\section{Related Work}
\label{sec:RelatedWork}

Efficient management of high-dimensional data underpins applications in recommendation,
multimedia retrieval, and large-scale biological and language modeling~\cite{10.1145/3448016.3457550}.
To meet the computational demands of similarity search at scale, vector databases and ANN systems
have developed multiple indexing paradigms—clustering-based, quantization-based, graph-based, and
hashing-based—yet persistent production challenges remain, including high index build cost, large
index footprints, and limited scalability under strict memory and latency budgets~\cite{pan2023surveyvectordatabasemanagement}.
These constraints complicate system deployment and tuning, while bounded compute and I/O budgets
limit candidate exploration, causing practical efficiency to degrade despite strong offline accuracy.

\paragraph{Quantization and IVF-based indexes (FAISS IVF--PQ / OPQ)}
FAISS provides a widely used family of inverted-file indexes (IVF, IVF--PQ, IVF--OPQ) with
SIMD-optimized kernels and runtime tuning via \texttt{ParameterSpace}
(e.g., \texttt{nprobe})~\cite{johnson2017billionscalesimilaritysearchgpus,ref40}.
IVF--PQ partitions the space using $k$-means and compresses residuals with product quantization,
supporting efficient lookup-table--based asymmetric distance computation (ADC)~\cite{5432202}.
OPQ learns a rotation to reduce quantization distortion at fixed code size~\cite{6678503}.
These quantization-based pipelines provide strong recall--latency trade-offs, but can suffer from
list imbalance and non-trivial training/build overheads at large scale. We also
adopt state-of-the-art prebuilt IVF--PQ configurations released for the BigANN benchmark/competition,
which reflect heavily tuned quantization baselines at billion scale~\cite{simhadri2022bigann,bigann_readme}.

\paragraph{Graph-based ANN (HNSW, disk-backed hybrids, and GPU graphs).}
Graph-based indexes such as HNSW achieve high recall through hierarchical routing and small-world
connectivity~\cite{27,33}. Other graph families include Voronoi-style routing structures (e.g., HVS)~\cite{34}
and degree-reduced traversable graphs (e.g., NSG) that aim to lower traversal cost while maintaining
navigability~\cite{CongFu}. For billion-scale deployments, disk-backed hybrids such as DiskANN and SPANN
combine in-memory structures with SSD-aware graph storage and pruning to balance recall, latency,
and I/O efficiency~\cite{NEURIPS2019_09853c7f,NEURIPS2021_299dc35e}. Recent GPU-accelerated graph search
(e.g., CAGRA) further improves throughput by exploiting parallel traversal and memory bandwidth on
modern devices~\cite{ootomo2024cagrahighlyparallelgraph}. Despite their effectiveness, graph construction
is computationally intensive and typically scales poorly with dataset size and dimensionality; moreover,
achieving competitive performance often requires substantial memory/SSD resources and careful tuning.

\paragraph{Hashing and binary-code retrieval (LSH, ITQ, and learned hashing).}
Hashing maps high-dimensional vectors into compact binary codes to enable fast Hamming-space filtering
and low-memory indexing~\cite{hash,Liong_2015_CVPR,Lai_2015}. Classical LSH provides probabilistic candidate generation, including SimHash~\cite{10.1145/509907.509965} and $p$-stable LSH for $\ell_p$ distances~\cite{10.1145/997817.997857}, building on the original LSH framework~\cite{10.1145/276698.276876,26,10.14778/3594512.3594527,31,35}. For an overview of near-optimal hashing results for ANN, see~\cite{10.1145/1327452.1327494}.

These methods are simple and parallelizable, but often require multi-table designs or aggressive probing to achieve high recall in high dimensions, inflating candidate sets and query latency. Multi-Probe LSH ranks likely perturbations to probe multiple nearby buckets within a table~\cite{10.5555/1325851.1325958}, and Multi-Index Hashing accelerates multi-probe enumeration in Hamming space via substring indexing~\cite{norouzi2014fastexactsearchhamming}. While related in spirit, we drive probing using learned logits with a calibrated margin $\delta$ and enforce a strict budget via gating and early stopping.

Data-dependent hashing methods, including Spectral Hashing and ITQ, learn compact codes that better align with local neighborhoods compared with random projections~\cite{NIPS2008_d58072be,inproceedings2011},
but are commonly evaluated at smaller scales than billion-vector benchmarks. FAISS also provides \texttt{IndexLSH}, a training-free SimHash implementation typically paired with exact re-ranking for competitive recall~\cite{ref40}. Learned hashing further explores neural encoders to produce binary codes; TBH is a representative learned hashing pipeline that combines neural encoding with compact code generation and serves as a strong learned baseline in our evaluation~\cite{tbh}.

\paragraph{Learned index structures and learned quantization.}
Learned index structures replace static indexing components (e.g., trees, hashes, quantizers) with
models that exploit data regularities. Kraska et al.~\cite{kraska2018case} introduced this viewpoint,
with follow-up systems such as ALEX~\cite{ding2020alex} and LISA~\cite{li2020lisa} demonstrating improved
adaptivity in ordered/spatial settings. In vector retrieval, a parallel
trend appears in learned quantization and discrete representation learning. Early work (e.g., DQN) jointly
optimized representation learning and quantization for retrieval~\cite{cao2016deepquantization}, while more
recent methods use distillation or constrained clustering to learn compact codes for dense retrieval
(e.g., Distill-VQ, contrastive distillation variants, RepCONC)~\cite{xiao2022distillvq,oneill2023improvedvq,zhan2023repconc}.

\paragraph{Position of NeuRoute.}
NeuRoute is most closely related to learned hashing and discrete representations, but it is designed for billion-scale routing: a lightweight training phase produces compact binary codes for Hamming-space navigation and efficient candidate generation. Compared to graph- and disk-backed methods, NeuRoute reduces dependency on complex graph maintenance and storage/I/O-specific tuning. Compared to IVF–PQ pipelines, it avoids global coarse clustering and heavyweight quantizer training, relying instead on bucketized indexing with  lightweight bucket-local clustering in low dimension. Overall, NeuRoute targets scalable retrieval with low index-structure overhead and fast time-to-index, while using exact refinement for final ranking.

\section{Methodology} \label{sec:Methodology}

This section presents the proposed \textit{NeuRoute} framework for large-scale similarity search in vector databases. We begin by formally defining the approximate $k$-nearest neighbor (ANN) search problem. We then detail the design of (1) a learnable \emph{hash function}—including the model architecture, loss formulation, and binarization process—and (2) the \emph{index construction and retrieval} mechanism. Finally, we provide a brief \emph{time-complexity} analysis.

\subsection{Problem Definition}

Let $\mathcal{X} = \{\mathbf{x}_i\}_{i=1}^N \subset \mathbb{R}^{E_{\text{dim}}}$ denote a set of $N$ input embeddings, where $\mathbf{x}_i$ represents the embedding of a sample in a high-dimensional space of dimension $E_{\text{dim}}$. Given a query $\mathbf{q} \in \mathbb{R}^{E_{\text{dim}}}$ and the Euclidean distance
$d(\mathbf{u},\mathbf{v})=\|\mathbf{u}-\mathbf{v}\|_2$,
the goal of approximate $k$-nearest neighbor search is to retrieve a subset

\begin{equation}
 S_k(\mathbf{q}) \subset \mathcal{X}, \quad |S_k(\mathbf{q})| = k,   
\end{equation}

such that the retrieved elements are close approximations of the true nearest neighbors:
\begin{equation}
S_k(\mathbf{q}) \approx \operatorname*{argmin}_{S: |S|=k} \sum_{\mathbf{x} \in S} d(\mathbf{q}, \mathbf{x}).
\end{equation}
To enable efficient routing-based retrieval at scale, \textit{NeuRoute} learns encoder logits that approximately preserve local neighborhood structure; the resulting short binary codes support fast multi-bucket probing, followed by centroid-based cluster filtering and exact refinement in the original embedding space.

Figure~\ref{fig:spoverview} provides a roadmap of NeuRoute (Steps~1--5). 
Steps~1--2 learn the hash function: embeddings are mapped to logits by the encoder, and then binarized by median thresholding to obtain balanced binary addresses. 
Step~3 builds the bucketized index and performs in-bucket clustering in the latent space, while storing original embeddings separately for exact refinement. 
At query time, Step~4 enumerates candidate buckets around the base address using a logit-guided procedure, and Step~5 selects candidate clusters via centroid distances with calibrated gating and heap-quality-driven early stopping before exact scoring.  

This interleaved routing–refinement pipeline yields bounded candidate sets, supports early termination when no further improvements occur, and achieves a well-balanced trade-off among retrieval speed, memory efficiency, and search accuracy.

\subsection{Hash Function}
\label{subsec:hash-function}

The \textit{NeuRoute} framework efficiently transforms high-dimensional input embeddings into compact binary codes for rapid similarity search. This process involves two key transformations: (1) mapping vectors from the embedding space to a low-dimensional latent space via a neural encoder, and (2) binarizing the latent representations to produce discrete addresses in the Hamming space.

\subsubsection{Encoder Definitions}
\label{subsubsec:encoder}

At the core of \textit{NeuRoute} lies an encoder architecture that compresses high-dimensional embeddings into compact latent representations and preserves their similarity. The encoder $E(\cdot)$  is formally defined as:
\begin{align}
 \boldsymbol{\ell} &= E(\mathbf{X}), \quad \boldsymbol{\ell} = [\boldsymbol{\ell}_1, \boldsymbol{\ell}_2, \dots, \boldsymbol{\ell}_N] \in \mathbb{R}^{N \times L_{\text{dim}}}
\end{align}
where $\mathbf{X} = [\mathbf{x}_1, \mathbf{x}_2, \dots, \mathbf{x}_N] \in \mathbb{R}^{N \times E_{\text{dim}}}$ denotes the input embeddings, $\boldsymbol{\ell}$ the latent representations. Here, $L_{\text{dim}} \ll E_{\text{dim}}$ represents the dimension of the latent space.

The encoder $E$ extracts the most salient features of the high-dimensional input while suppressing redundancy. 

\subsubsection{Loss Function Design}
\label{lossfunction}

The \textit{NeuRoute} framework employs a selective similarity-preservation loss. Preserving the complete similarity structure is difficult when the number of samples is large, while ANN search only requires preserving relationships among nearby vectors. We therefore propose the selective similarity-preservation loss $L_{\text{sim}}$.

The selective similarity loss ensures that neighborhood relationships in the embedding space are preserved in the latent space.
For a mini-batch of $B$ vectors, let $\rho=\sqrt{L_{\text{dim}}/E_{\text{dim}}}$ denote the dimensionality scaling factor, and define two pairwise distance matrices under Euclidean distance:
$\mathbf{S}_{\text{emb}} \in \mathbb{R}^{B \times B}$ for the original embeddings and
$\mathbf{S}_{\text{lat}} \in \mathbb{R}^{B \times B}$ for the latent representations.
The loss measures their discrepancy over selected pairs, using a masking mechanism to focus on semantically meaningful relationships:
\begin{equation}
\label{similarityconfig}
L_{\text{sim}} = \frac{1}{\mathrm{sum}(\mathrm{Mask})} 
\sum_{i=1}^{B} \sum_{j=1}^{B}
\mathrm{Mask}_{ij}
\Big(
  \mathbf{S}_{\text{lat}}(i,j)
  - \gamma\rho \cdot
  \mathbf{S}_{\text{emb}}(i,j)
\Big)^{2},
\end{equation}
where $\mathrm{sum}(\cdot)$ denotes the element-wise summation of the mask matrix and $\gamma$ is a scaling coefficient.
The pairwise distance matrices are defined as:
\begin{align}
\mathbf{S}_{\text{emb}}(i,j) &= \|\mathbf{x}_i - \mathbf{x}_j\|_2, \\
\mathbf{S}_{\text{lat}}(i,j) &= \|\boldsymbol{\ell}_i - \boldsymbol{\ell}_j\|_2.
\end{align}

\paragraph{Metric.}
Throughout this paper, we use Euclidean distance:
\begin{equation}
d(x,y)=\|x-y\|_2.
\end{equation}

The binary mask $\mathrm{Mask} \in \{0,1\}^{B \times B}$ identifies significant pairs based on similarity thresholds:
\begin{equation}
\label{equation8}
\mathrm{Mask}(i,j)=
\begin{cases}
1, & i\ne j\ \text{and}\ \bigl(\mathbf{S}_{\text{emb}}(i,j)\le \tau_{\text{emb}}
\ \ \text{or}\ \
\mathbf{S}_{\text{lat}}(i,j)\le \gamma\rho\,\tau_{\text{emb}}\bigr), \\[4pt]
0, & \text{otherwise.}
\end{cases}
\end{equation}
Here, $\tau_{\text{emb}}$ defines the similarity threshold controlling the model’s sensitivity to meaningful relationships. This selective focus enables \textit{NeuRoute} to emphasize informative embedding pairs while ignoring weakly correlated ones.

Our objective is not to preserve the global similarity structure, but to optimize for ANN: we only need the very most similar pairs to remain reliably similar. Therefore, $L_{\text{sim}}$ enforces consistency between the latent similarity $S_{\text{lat}}$ and the input-space similarity $S_{\text{emb}}$ only on top-quantile (nearest-neighbor) pairs (the mask selects the most similar $\sim 0.5\%$), concentrating model capacity on the region that directly determines recall. In addition, the dual-threshold anti-collapse mechanism prevents non-neighbors from becoming ``artificially similar'' in latent space, which helps avoid bucket explosion and unnecessary candidates. The scatter plots show a clear monotonic alignment in the near-neighbor regime while remaining well-behaved outside it, indicating that the objective is directly aligned with ANN's core trade-off: high recall with a controlled candidate set size.

This loss also eliminates expensive data preparation (e.g., building large-scale pair lists, mining hard negatives, or running offline neighbor graphs): the mask and thresholds are computed on-the-fly from within-batch similarities. As a result, training integrates naturally into the pipeline and reduces preprocessing overhead, which is a key enabler for our $\sim$1-hour end-to-end indexing workflow.

\subsubsection{Binarization}
\label{subsubsec:binarization}

For each latent dimension $j$, a per-dimension threshold $\tau_j$ is defined as the median activation across all samples:
\[
\tau_j = \operatorname{median}_{i=1,\dots,N} \, \ell_i[j], 
\qquad j = 1,\dots,L_{\text{dim}}.
\]
Each latent activation is then binarized independently:
\begin{equation}
\label{equation10}
a_i[j] =
\begin{cases}
1, & \ell_i[j] \ge \tau_j, \\[3pt]
0, & \ell_i[j] < \tau_j,
\end{cases}
\qquad i = 1,\dots,N.
\end{equation}
Here, $\ell_{i}[j]$ denotes the $j$-th component of the latent vector for sample $i$, and $a_i[j]$ represents the corresponding binary bit.
This median-threshold binarization encourages balanced bit distributions across the dataset, stabilizing hash-bucket allocation.
In practice, we estimate the per-dimension median thresholds on the same subsampled base used for the bucket-shape check, incurring no additional data pass; the time breakdown is reported in Table~\ref{tab:NeuRoute_time_breakdown}.

\paragraph{Deviation scores for bucket enumeration.}
After binarization, we define a per-dimension deviation score for a query $q$ that measures how far the activation is from the threshold:
\begin{equation}
\delta_q[j] = \left| \ell_q[j] - \tau_j \right|,\quad j=1,\ldots,L_{\text{dim}}.
\end{equation}
Intuitively, a larger $\delta_q[j]$ indicates a more stable bit decision (farther from the threshold), while a smaller $\delta_q[j]$ suggests higher uncertainty near the threshold. We use $\{\delta_q[j]\}$ to select informative dimensions (e.g., the $L_{\text{sel}}$ least-stable dimensions with the smallest $\delta_q[j]$) and to assign bit-flip costs when enumerating candidate buckets around the query's base binary address (Section~\ref{subsec:retrieval}).

\subsubsection{Distance Preservation Diagnostics}

\begin{figure}[t]
  \centering
  \includegraphics[width=0.80\textwidth]{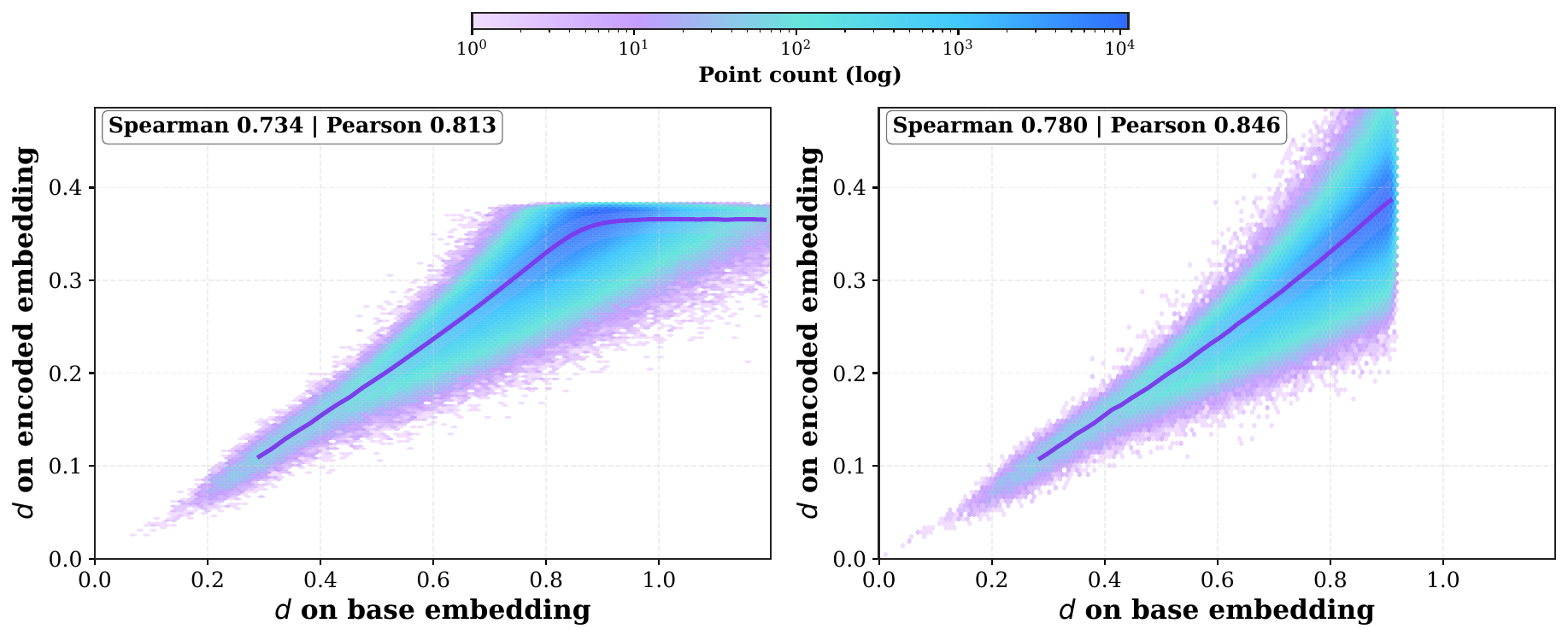}
  \caption{Pairwise distances in the base space versus the encoded space for paired vectors
  (squared for visualization; squaring is monotonic for nonnegative distances and does not change top-$k$ ordering).
  Left: encoded-sampled pairs. Right: base-sampled pairs.}
  \label{fig:d_base_vs_encoded_2d}
\end{figure}

To better understand how the learned encoding reshapes geometry, we visualize the relationship between pairwise distances
measured in the original space and in the encoded space.
Figure~\ref{fig:d_base_vs_encoded_2d} shows two panels based on pre-sampled paired vectors:
pairs sampled via the encoded space (left) and pairs sampled via the base space (right).
Across both sampling schemes, the encoded distances remain strongly monotonic with the base distances, while large-distance pairs
exhibit mild saturation in the encoded space. This behavior is consistent with our objective of preserving relative similarity for near
neighbors, and it supports stable heap-quality signals for early stopping.

\subsection{Index Construction}
\label{subsec:index-build}

Figure~\ref{fig:spoverview} illustrates index construction. We start from the embedding space, where embeddings
$\mathbf{X}\in\mathbb{R}^{N\times E_{\text{dim}}}$ are mapped by the encoder $E(\cdot)$ into a compact latent representation
$\boldsymbol{\ell}=E(\mathbf{X})\in\mathbb{R}^{N\times L_{\text{dim}}}$.
Each latent vector is then binarized into a short address
$\mathbf{A}\in\{0,1\}^{N\times L_{\text{dim}}}$, enabling bucketized indexing in Hamming space.
\paragraph{Bucket allocation.}
For each vector $x_i$, its binary address $a_i$ determines the target bucket $b(a_i)$.
We append the identifier $i$ to the posting list of $b(a_i)$, and keep only non-empty buckets in the directory.
The original embeddings $\{x_i\}$ are stored separately for exact top-$k$ refinement.

\paragraph{In-bucket clustering in latent space.}
For each non-empty bucket $b$ (i.e., the bucket indexed by an address $u$ with a non-empty posting list), we collect the latent representations
$\{\ell_i \mid i \in b\}$ of all vectors assigned to this bucket and perform clustering within the bucket in the latent space.
We adopt a size-aware strategy based on bucket cardinality: very small buckets are not clustered and are treated as a single cluster,
while medium and large buckets use different target cluster capacities.
For a bucket $B$ (with latent set $\{\ell_i \mid i\in B\}$), the number of clusters is determined adaptively as
\[
K_B=\left\lceil \frac{|B|}{t} \right\rceil,
\]
where $t$ denotes the target cluster capacity.
We first run k-means with $K_B$ clusters on a sampled subset of $\{\ell_i\}$ with k-means++ initialization to obtain initial centroids,
and then apply a limited number of Lloyd refinement passes on the full set $\{\ell_i \mid i \in B\}$ in that bucket.
After clustering, each vector is assigned a local cluster label within its bucket, and a centroid representation is produced for each local cluster.

\subsection{Retrieval Process}
\label{subsec:retrieval}

As shown in Figure~\ref{fig:spoverview} (Steps~4--5), NeuRoute performs \emph{budgeted, prioritized} candidate generation before exact scoring. Given a query $q$, we first enumerate promising binary addresses (buckets) around its hash address under a bounded exploration budget (Step~4). We then process the resulting buckets in that order, evaluating their bucket-local cluster centroids and maintaining a bounded candidate set using calibrated distance gating and heap-quality-driven early stopping (Step~5). Finally, we expand the retained clusters into vector candidates and score them exactly in the original embedding space.

We use $d_{\text{cent}}(\cdot,\cdot)$ to denote the centroid-level distance used for candidate filtering, while exact scoring uses Euclidean distance $d(\cdot,\cdot)$ in the original embedding space, where $d(x,y)=\|x-y\|_2$.

\begin{algorithm}[t]
\caption{Bucket Enumeration around the Base Address}
\label{alg:bucket_enum}
\begin{algorithmic}[1]
\Require
Query logit vector $\boldsymbol{\ell}_q \in \mathbb{R}^{L_{\text{dim}}}$;
thresholds $\boldsymbol{\tau}\in\mathbb{R}^{L_{\text{dim}}}$;
selection size $L_{\text{sel}}$;
max flip radius $R_{\max}$;
number of score bins $N_{\text{bins}}$;
filter: max\_score $S_{\max}$.
\Ensure
Ordered bucket bins $\mathcal{B}_q[0..N_{\text{bins}}-1]$ (each is a list of bucket IDs).
 
\State $a_q[j] \gets \mathbb{I}\!\left[\ell_q[j] \ge \tau_j\right]\;\;\;\forall j=1,\dots,L_{\text{dim}}$
\State $b_0 \gets \textsc{BucketId}(a_q)$
\State $\delta_q[j] \gets \left|\ell_q[j]-\tau_j\right|\;\;\;\forall j=1,\dots,L_{\text{dim}}$
\State $\mathcal{S} \gets \textsc{SmallestKIndices}(\{\delta_q[j]\}_{j=1}^{L_{\text{dim}}},\, L_{\text{sel}})$
 
\State Initialize empty bins $\mathcal{B}_q[0..N_{\text{bins}}-1]$
\State Append $b_0$ to $\mathcal{B}_q[0]$ \Comment{score $s=0$}
 
\ForAll{$\Delta \subseteq \mathcal{S}$ with $1\le|\Delta|\le R_{\max}$}
    \State $b \gets \textsc{ApplyFlips}(b_0, \Delta)$
    \State $s \gets \sum_{j\in\Delta}\delta_q[j]$
    \If{$s \le S_{\max}$}
        \State $z \gets \textsc{Bin}(s; N_{\text{bins}}, S_{\max})$
        \State Append $b$ to $\mathcal{B}_q[z]$
    \EndIf
\EndFor
\State \Return $\mathcal{B}_q$
\end{algorithmic}
\end{algorithm}

In our implementation, we bootstrap the calibration top-$K$ lists by running the same end-to-end pipeline on a held-out set of calibration queries under a higher-recall setting (i.e., with an enlarged search budget/range), and use the resulting top-$K$ outputs as supervision for calibration.

\paragraph{Bucket-score cutoff ($S_{\max}$).}
For each $(q,\text{target})$ pair in the top-$K$ list, we compute the target bucket score using the same scoring function as in bucket enumeration.
We then measure coverage as a function of a candidate threshold $t$: (i) the fraction of top-$K$ pairs with score $\le t$, and (ii) the fraction of queries whose entire top-$K$ set is covered.
We choose $S_{\max}$ to achieve high top-$K$ bucket coverage while keeping the enumeration cost bounded.

\paragraph{Centroid gating margin ($\mathcal{M}$).}
For each query, we scan all centroids within the buckets that contain its top-$K$ targets and record (1) the best-seen centroid distance $x$ and (2) the additional margin $y$ required so that the farthest top-$K$ target would still pass a distance gate.
We estimate a data-driven margin curve $\mathcal{M}(x)$ by grouping queries into equal-frequency (quantile) bins by $x$ and taking a high quantile (e.g., 95th or 99th) of $y$ within each bin, smoothing the result into a monotone piecewise-linear function.
At inference time, the centroid-stage gate applies the threshold $\theta(x)=x+\mathcal{M}(x)$.

\subsubsection{Bucket Enumeration (Step~4)}
\label{subsubsec:dp-enum}

Algorithm~\ref{alg:bucket_enum} summarizes Step~4. Given the query hash address $a_q=h(q)$, we apply an enumerator to generate and rank nearby binary addresses under a bounded enumeration budget. Enumerated addresses are grouped into priority bins and processed from high-yield to low-yield bins, enabling termination once additional enumeration no longer improves the retained candidates.

\subsubsection{Centroid-Stage Selection with Calibrated Gating and Early Stop (Step~5)}
\label{subsubsec:centroid-stage}

We process the enumerated buckets in increasing bin-score order (Alg.~1).
Buckets that contain a single bucket-local cluster are added directly to $C^{\mathrm{dir}}(q)$,
since centroid-level selection offers limited benefit in this case.

For the remaining (non-singleton) buckets, we expand them to their bucket-local centroids and perform centroid-stage selection as summarized in Alg.~\ref{alg:centroid_enum_gate_simpleB}: we compute centroid distances in batches, maintain a bounded top-$K_c$ heap, and apply a calibrated gate $\theta = d_{\min} + M(d_{\min})$ to suppress low-value candidates. Early stopping is enabled only after the heap is full and a configurable minimum number of bins $B_{\min}$ has been processed. For each centroid batch, we count qualifying heap updates and, using the front threshold $d_{\min}+\alpha_{\mathrm{front}}\mathcal{M}(d_{\min})$, updates that improve the heap front. A batch is low quality when these update counts fall below the configured thresholds; we stop after $T_{\mathrm{low}}$ consecutive low-quality batches.

The centroid stage returns $C^{\mathrm{cent}}(q)$ (clusters retained in the heap), and the pre-refinement
candidate set is
\[
C(q) = C^{\mathrm{dir}}(q)\ \cup\ C^{\mathrm{cent}}(q).
\]

\paragraph{Implementation.}
We implement the end-to-end retrieval pipeline in C++ for efficiency and reproducibility. The index is stored in a CSR-style layout to represent variable-length posting lists (e.g., bucket- and cluster-level inverted structures) compactly and enable sequential scans. At query time, we run the neural encoder via TorchScript (C++ API) to produce logits, and execute bucket/cluster enumeration and subsequent scanning entirely in C++ using vectorized distance kernels and multi-threading. Bucket enumeration is implemented with a DP-based generator to efficiently enumerate candidate addresses under the configured radius/score budget.

\begin{algorithm}[t]
\caption{Centroid-Stage Cluster Enumeration with Calibrated Gating}
\label{alg:centroid_enum_gate_simpleB}
\small
\begin{algorithmic}[1]
\Require
Query logits $\boldsymbol{\ell}_q$;
ordered bucket bins $\mathcal{B}_q$ (Alg.~\ref{alg:bucket_enum});
bucket-to-cluster map $\Phi$; centroids $\{\boldsymbol{\mu}_c\}$;
calibration margin $\mathcal{M}(\cdot)$;
heap size $K_c$; batch cap $B_{\text{batch}}$; early-stop parameters $\alpha_{\mathrm{front}}$, $B_{\min}$, and $T_{\mathrm{low}}$.
\Ensure Candidate cluster set $C(q)$.
 
\State Initialize top-$K_c$ heap $H$, $d_{\min}\gets+\infty$, $b_{\mathrm{proc}}\gets0$, and $t_{\mathrm{low}}\gets0$
\For{$z$ in increasing order of bin score}
    \State $b_{\mathrm{proc}}\gets b_{\mathrm{proc}}+1$
    \State $C_z \gets \textsc{MapBucketsToClusters}(\mathcal{B}_q[z], \Phi)$
    \ForAll{batches $C_{batch}\subseteq C_z$ with $|C_{batch}|\le B_{\text{batch}}$}
        \State $\mathbf{d} \gets \textsc{CentroidDistances}(\boldsymbol{\ell}_q,\{\boldsymbol{\mu}_c:c\in C_{batch}\})$
        \State $d_{\min}\gets \min(d_{\min},\min(\mathbf{d}))$
        \State $\theta \gets d_{\min} + \mathcal{M}(d_{\min})$ \Comment{calibrated gate}
        \State \textsc{KeepTopKUnderGate}$(H, C_{batch}, \mathbf{d}, \theta)$
        \State Update $t_{\mathrm{low}}$ from heap/front updates in $C_{batch}$
        \If{$H$ is full, $b_{\mathrm{proc}}\ge B_{\min}$, and $t_{\mathrm{low}}\ge T_{\mathrm{low}}$}
            \State \textbf{break}
        \EndIf
    \EndFor
    \If{$H$ is full, $b_{\mathrm{proc}}\ge B_{\min}$, and $t_{\mathrm{low}}\ge T_{\mathrm{low}}$}
        \State \textbf{break}
    \EndIf
\EndFor
\State \Return $C(q)\gets \{c \mid (d,c)\in H\}$
\end{algorithmic}
\end{algorithm}

\subsection{System Complexity and Scalability Analysis}
This section gives a stage-level characterization rather than a tight worst-case bound, since runtime is governed by adaptive budgets.
For a query $q$, let $B_q$ be the number of enumerated bucket addresses, $C_q$ the number of visited centroids in latent space, and $R_q$ the number of candidates entering exact refinement.
The dominant query-time costs are centroid scoring in the latent space, approximately $O(C_q \cdot L_{\text{dim}})$, and refinement in the original embedding space, approximately $O(R_q \cdot E_{\text{dim}})$ under (squared) Euclidean distance.
In practice, $B_q$, $C_q$, and $R_q$ are controlled by the enumeration budget, calibrated gating, and heap-quality-driven early stopping.

For index construction, hash bucket assignment is linear in the database size $N$.
Bucket-local clustering decomposes across buckets and is performed in the encoder's low-dimensional space, making it substantially cheaper than clustering in the original embedding dimension.
Its practical cost is governed by (i) the latent dimension $L_{\text{dim}}$, (ii) the bucket size distribution (work per bucket), (iii) the per-bucket cluster budget, and (iv) the number of clustering iterations.
Because the learned codes yield relatively balanced buckets, per-bucket workloads remain small and highly parallelizable; in our 1B-scale setting, we build centroids for roughly 4 million buckets in about 10 minutes as part of the overall indexing pipeline.

Overall, NeuRoute combines learned hashing with budget-aware candidate generation: bucket enumeration generates candidates, centroid-stage gating prunes low-yield regions, and early stopping limits work before exact refinement, enabling controllable per-query budgets at scale.

\section{Experiment} 
\label{sec:experimental_analysis}

This section presents an evaluation of the proposed \textit{NeuRoute} framework under realistic vector-database conditions, including high-dimensional embeddings and large-scale corpora.

\subsection{Datasets}
\label{sec:datasets}

\begin{table}[t]
\centering
\small
\caption{Overview of datasets used in our evaluation.}
\label{tab:datasets_all}
\begin{adjustbox}{max width=\linewidth}
\begin{tabular}{llll}
\toprule
\textbf{Dataset} & \textbf{Domain} & \textbf{Type} & \textbf{Size} \\
\midrule
\multicolumn{4}{l}{\emph{High-dimensional, moderate-scale evaluation sets}}\\
GLDv2 (aug.) \cite{ref38} & Landmark Recognition & Image & 9.39M \\
\midrule
\multicolumn{4}{l}{\emph{Moderate-scale subsets (for build/query baselines)}}\\
BigANN-100M \cite{bigann_readme} & SIFT Features & Image & 100M \\
Deep1B-100M \cite{7780595} & CNN Features & Image & 100M \\
\midrule
\multicolumn{4}{l}{\emph{Billion-scale, moderate-dimensional evaluation sets}}\\
BigANN-1B \cite{bigann_readme} & SIFT Features & Image & 1B \\
Deep1B-1B \cite{7780595} & CNN Features & Image & 1B \\
\bottomrule
\end{tabular}
\end{adjustbox}
\end{table}

We evaluate \textit{NeuRoute} in two regimes (i) high-dimensional, moderate-scale datasets and (ii) billion-scale, moderate-dimensional benchmarks—summarized in Table~\ref{tab:datasets_all}.

\paragraph{High-dimensional, moderate-scale.}
We evaluate a high-dimensional regime on GLDv2 using 1536D image embeddings (DINOv2).
This setting tests robustness under high dimensionality and a multi-million-scale corpus, complementing the 96D/128D billion-scale benchmarks.

\paragraph{Billion-scale.}
For the moderate-dimensional billion-scale benchmarks, we use BigANN-1B (hand-crafted SIFT; \(128\)-D; \(1\)B) and Deep1B-1B (CNN features; \(96\)-D; \(1\)B). Both come from the NeurIPS 2021 BigANN challenge \cite{simhadri2022bigann,bigann_readme}. We follow the prescribed base/query splits for all evaluations (Table~\ref{tab:datasets_all}).

\subsubsection{Dataset Preprocessing}  

\paragraph{Data preparation and embeddings.}
We expand the effective size of the Google Landmarks v2 (GLDv2) \emph{train}, \emph{index}, and \emph{test} splits with standard augmentations—random cropping, horizontal flipping, and color jitter—without changing the evaluation protocol. The augmented GLDv2 images are embedded with the \emph{DINOv2 (giant)}~\cite{ref43} model. We $\ell_2$-normalize both base and query embeddings before ground-truth computation and indexing. For unit-normalized vectors, cosine similarity, inner product, and squared Euclidean distance induce the same neighbor ordering; we report Euclidean distance for consistency with the other datasets.

\paragraph{BigANN specifics.}
BigANN vectors are stored as integers (typically \texttt{uint8}). To make them compatible with neural training and index construction, we apply a byte-scale normalization by dividing by 255 \emph{only during encoder training}. 
For evaluation and ground-truth computation, \(k\)-NN search and distance calculations are performed on the \emph{original} integer database. 
This preserves comparability with FAISS baselines while enabling stable model fitting on normalized inputs.

\subsection{Experiment Environment Setup}
Our experiments were conducted on a single host with an \textbf{AMD EPYC 7543} CPU
(2 sockets $\times$ 32 physical cores; 64 physical cores total; SMT enabled with 128 hardware threads)
and 1007\,GiB RAM across 8 NUMA nodes ($\sim$126\,GiB per node), together with one
\textbf{NVIDIA A100-PCIE-40GB} GPU.
All datasets and index artifacts were stored on a \textbf{Lustre} parallel file system mounted at \textbf{/scrfs};
the host also provides \textbf{1.6\,TB} of local NVMe SSD, which is used when required by disk-backed baselines
(e.g., DiskANN).

The GPU is used only for training learned hashing models (NeuRoute and TBH) using
\textbf{PyTorch~1.13.1} (Python~3.10, Linux~x86\_64).
Unless otherwise stated, all reported encoding, index construction, query-time retrieval,
and evaluation are executed on the CPU.

For CPU baselines, we use a consistent multithreading setting of 24 threads whenever applicable
(e.g., FAISS OpenMP threads, HNSW search threads, and DiskANN's $T$ parameter),
and we explicitly note any single-thread measurements or method-specific constraints.

\subsection{Baselines}
\label{subsec:baselines}

\textbf{Baseline selection and evaluation knobs.}
We evaluate ANN baselines that are reproducible and feasible on a single node up to 1B scale.
To keep comparisons interpretable, we sweep one primary query-time knob per family to produce Recall@10--QPS frontiers:
\texttt{nprobe} for IVF-based indexes, \texttt{efSearch} for HNSW, search depth $L$ for DiskANN, and a single probing-budget scalar for 22-bit hashing ($b=22$).
Fixed build/training configurations are summarized in Table~\ref{tab:baseline_config}.
At 1B scale, methods requiring full-resident vectors or large in-memory graphs (e.g., IVF-Flat and HNSW) are omitted under our single-node memory/storage/wall-time budgets; instead, we report OPQ+IVF-PQ with bounded refinement as an accuracy-oriented, disk-feasible reference.

\noindent\textbf{Baselines.}
\textbf{Graph (CPU).} We include FAISS \texttt{IndexHNSWFlat} with fixed build settings ($M{=}32$, \texttt{efConstruction}$=200$) and sweep \texttt{efSearch} on GLDv2 and the 100M subsets.
\textbf{Graph (GPU reference).} We additionally report CAGRA~\cite{ootomo2024cagrahighlyparallelgraph} on 80M subsets due to the 40\,GB device-memory limit, treated as a GPU-only throughput reference.
\textbf{Disk-based.} We include DiskANN~\cite{NEURIPS2019_09853c7f} using the official implementation on local NVMe and sweep the search depth $L$.
\textbf{Quantization.} We include FAISS IVF-Flat/IVF-PQ~\cite{johnson2017billionscalesimilaritysearchgpus,5432202} where feasible, train codebooks on a fixed subset, build by streaming adds, and sweep \texttt{nprobe}.
\textbf{1B refinement reference.} On 1B, we also report OPQ+IVF-PQ with bounded exact reranking~\cite{6678503,5432202} by sweeping $k_{\text{factor}}\in\{50,100,200\}$ for $k=10$ and reporting end-to-end query time.
\textbf{Hashing.} We include 22-bit hashing baselines (LSH, ITQ~\cite{itq}, TBH~\cite{tbh}) and sweep a single probing-budget scalar; we additionally report bucket-occupancy and margin--flip diagnostics.

\begin{table}[t]
\centering
\footnotesize
\setlength{\tabcolsep}{3pt}
\renewcommand{\arraystretch}{1.02}
\caption{Configurations used in our evaluation. Baselines use fixed settings. NeuRoute lists query-time routing parameters (ranges indicate multiple configs observed in logs).}
\label{tab:baseline_config}
\begin{adjustbox}{max width=\linewidth}
\begin{tabular}{@{}llp{0.68\linewidth}@{}}
\toprule
Setting & Method & Configuration \\
\midrule
GLDv2 (1536D) & IVF-Flat & \texttt{IVF49152} \\
GLDv2 (1536D) & IVF-PQ & \texttt{OPQ96,IVF49152,PQ96x8} \\
\midrule
BigANN-100M (128D) & IVF-PQ & \texttt{IVF49152,PQ32x8} \\
Deep1B-100M (96D) & IVF-Flat & \texttt{IVF49152} \\
Deep1B-100M (96D) & IVF-PQ & \texttt{IVF49152,PQ24x8} \\
\midrule
BigANN-1B (128D) & IVF-PQ & \texttt{IVF49152,PQ32x8} \\
Deep1B-1B (96D) & IVF-PQ & \texttt{IVF49152,PQ24x8} \\
BigANN-1B (128D) & OPQ+IVF-PQ (refine) & \texttt{OPQ,IVF65536,PQ32x8} \\
Deep1B-1B (96D) & OPQ+IVF-PQ (refine) & \texttt{OPQ,IVF65536,PQ16x8} \\
\midrule
As available & HNSW & \texttt{IndexHNSWFlat} ($M{=}32$, \texttt{efC}{=}200) \\
GLDv2 (1536D) & DiskANN & $R{=}90,\ L_{\mathrm{build}}{=}100,\ B{=}50,\ M{=}200,\ T{=}24$ \\
BigANN-100M & DiskANN & $R{=}100,\ L_{\mathrm{build}}{=}100,\ B{=}50,\ M{=}80,\ T{=}24$ \\
Deep1B-100M & DiskANN & $R{=}100,\ L_{\mathrm{build}}{=}100,\ B{=}50,\ M{=}110,\ T{=}24$ \\
BigANN-1B & DiskANN & $R{=}48,\ L_{\mathrm{build}}{=}64,\ B{=}50,\ M{=}80,\ T{=}24$ \\
Deep1B-1B & DiskANN & $R{=}90,\ L_{\mathrm{build}}{=}100,\ B{=}20,\ M{=}300,\ T{=}24$ \\
As available & CAGRA (GPU) & degree=32, inter=64, itopk=128, iter=100 \\

As available & LSH / ITQ / TBH & 22-bit codes \\
\midrule
GLDv2 (1536D) & NeuRoute &
\(L_{\mathrm{sel}}{=}12\text{--}13,\ R_{\max}{=}4\text{--}5,\ K_c{=}10\text{--}1000,\ S_{\max}{=}0.07\text{--}0.09\) \\
BigANN-100M (128D) & NeuRoute &
\(L_{\mathrm{sel}}{=}14\text{--}16,\ R_{\max}{=}5\text{--}6,\ K_c{=}20\text{--}5000,\ S_{\max}{=}0.24\text{--}0.41\) \\
Deep1B-100M (96D) & NeuRoute &
\(L_{\mathrm{sel}}{=}14\text{--}15,\ R_{\max}{=}5\text{--}7,\ K_c{=}10\text{--}5000,\ S_{\max}{=}0.25\text{--}0.39\) \\
BigANN-1B (128D) & NeuRoute &
\(L_{\mathrm{sel}}{=}16,\ R_{\max}{=}6\text{--}7,\ K_c{=}500\text{--}5000,\ S_{\max}{=}0.36\) \\
Deep1B-1B (96D) & NeuRoute &
\(L_{\mathrm{sel}}{=}16\text{--}17,\ R_{\max}{=}4\text{--}7,\ K_c{=}500\text{--}10000,\ S_{\max}{=}0.25\text{--}0.36\) \\
\bottomrule
\end{tabular}
\end{adjustbox}
\vspace{-2mm}
\end{table}

\subsection{Metrics}\label{sec:metrics}

\paragraph{Recall.}
We evaluate retrieval completeness using \emph{Recall@\(k\)} with relevance judged \emph{only by unique IDs}.
For query \(i\), let \(\mathcal{G}_i^{k}\) be the ground-truth top-\(k\) ID set obtained by exact search (excluding the query itself), and let \(\mathcal{R}_i^{k}\) be the system's returned top-\(k\) ID set. The per-query recall is
\begin{equation}
r_i(k) \;=\; \frac{\bigl|\mathcal{R}_i^{k}\cap \mathcal{G}_i^{k}\bigr|}{\min\!\bigl(k,\;|\mathcal{G}_i^{k}|\bigr)}\, .
\end{equation}
Averaging over \(n_q\) queries yields
\begin{equation}
\mathrm{Recall@}k \;=\; \frac{1}{n_q}\sum_{i=1}^{n_q} r_i(k)\, .
\end{equation}

\paragraph{Quality–Throughput}
Throughout, we report \textbf{Recall@\(k\) versus queries per second (QPS)} \cite{ref40} to show the quality--throughput trade-off, where
\(\mathrm{QPS} = n_q / T_{\text{wall}}\) (total number of queries divided by wall-clock time); in figures,
points farther up and to the right are better. For a hardware-agnostic view, we also report and plot
\textbf{Recall@\(k\) vs.\ average candidates per query}—the mean number of database vectors
that receive an exact distance computation during refinement (labeled \emph{“average candidates”}
in the figures). This quantity is proportional to the amount of work and is comparable across systems.

\subsection{Experimental Setup and Design}
\label{experiment_design}

We design evaluations to reflect practical large-scale vector search settings rather than toy benchmarks. Unless otherwise noted, we use Euclidean distance and fix $k{=}10$ across all experiments, matching common top-$k$ retrieval usage in recommendation systems, document search, and RAG-style pipelines \cite{rajput2023recommendersystemsgenerativeretrieval,wang2018billionscalecommodityembeddingecommerce,10.1145/3448016.3457550,ref3,ref28}. We evaluate on high-dimensional embeddings and large-scale corpora (up to billion-entry bases) under strict runtime budgets in a single-node setup.

\paragraph{Benchmarks.}
We evaluate NeuRoute on (i) billion-scale benchmarks: BigANN-1B and Deep1B-1B; (ii) medium-scale counterparts: BigANN-100M and Deep1B-100M; and (iii) a high-dimensional robustness benchmark: GLDv2 (10M, 1536-d). Unless otherwise stated, each dataset uses $n_q=10{,}000$ queries, and all methods are evaluated in the same batched setting. Configurations for all baselines and NeuRoute are summarized in Table~\ref{tab:baseline_config}; unless stated otherwise, NeuRoute uses $\alpha_{\mathrm{front}}=0.5\text{--}0.7$ for the heap-quality early-stopping trigger.

\subsubsection{NeuRoute Model and Hyperparameters}
\label{sec:neuroute_model_hparams}

\paragraph{Architecture.}
NeuRoute uses a lightweight MLP encoder that progressively compresses the input embedding from $E_{\text{dim}}$ to a compact latent logit vector of dimension $L_{\text{dim}}$. Each hidden layer follows Linear$\rightarrow$BatchNorm$\rightarrow$ReLU, while the final encoder layer is linear to preserve the latent logits used for binarization and routing. Model instantiations differ only in layer widths to match dataset dimensionality; the layer pattern and training objective remain the same. At retrieval time, NeuRoute runs a single forward pass of the encoder $E(\cdot)$ to obtain the latent vector.

\paragraph{Choosing $L_{\text{dim}}$ (bucket-count trade-off).}
We choose the code length $L_{\text{dim}}$ to control the number of buckets and balance two competing costs. If $L_{\text{dim}}$ is too large, the address space becomes overly sparse and query-time bucket enumeration becomes expensive (e.g., more buckets to probe and/or larger effective radii to reach sufficient candidates). If $L_{\text{dim}}$ is too small, buckets become too coarse, increasing per-bucket workload and making bucket-local clustering and refinement scans more expensive. In practice, we choose $L_{\text{dim}}$ based on dataset scale to keep bucket sizes manageable while maintaining an affordable enumeration budget. In our experiments, we use $L_{\text{dim}}{=}22$ for the 1B-scale benchmarks, $L_{\text{dim}}{=}20$ for the 100M-scale setting, and $L_{\text{dim}}{=}16$ for GLDv2. The isolated pure-hashing comparison in Section~\ref{subsec:hashing_context} uses 22-bit codes for all methods.

\paragraph{Training protocol.}
We train NeuRoute using a fixed 2M-vector subset per dataset. We use mini-batches of size 4096 and optimize with Adam (learning rate $2\times 10^{-4}$) for 500 epochs on all datasets. Under this schedule, training typically stabilizes early; full training and validation curves are provided in Appendix~\ref{app:train_val_loss}.

\paragraph{Objective and fixed hyperparameters.}
We optimize the selective similarity-preservation loss $L_{\text{sim}}$, which aligns local neighborhood structure between the original embedding space and the latent space. We fix the scaling coefficient $\gamma=0.6$ across datasets, and additionally apply a dimensionality-reduction factor $\sqrt{L_{\mathrm{dim}}/E_{\mathrm{dim}}}$; i.e., the overall scaling is $\gamma \sqrt{L_{\mathrm{dim}}/E_{\mathrm{dim}}}$.

\paragraph{Selecting $\tau_{\text{emb}}$.}
We choose the threshold $\tau_{\text{emb}}$ in a dataset-specific manner via batch subsampling. Concretely, within each mini-batch we compute the distribution of pairwise distances and set $\tau_{\text{emb}}$ so that only the closest 0.5\% pairs satisfy $S_{\text{emb}}(i,j)\le \tau_{\text{emb}}$, focusing the similarity constraint on retrieval-relevant neighborhoods. The resulting $\tau_{\text{emb}}$ values are reported in Appendix~\ref{app:encoder_config}.

\begin{table}[t]
\caption{NeuRoute build-time breakdown (seconds). Add/Build $=$ Index build $+$ CSR build $+$ Clustering $+$ Other/I/O, and Total $=$ Train $+$ Add/Build. Index build performs full-base encoding. CSR build constructs CSR files. Clustering is bucket-local clustering. Other/I/O is residual overhead.}
\label{tab:NeuRoute_time_breakdown}
\centering
\scriptsize
\setlength{\tabcolsep}{3pt}
\renewcommand{\arraystretch}{1.05}
\resizebox{\textwidth}{!}{%
\begin{tabular}{l r r r r r r r}
\toprule
Dataset & Train (s) [GPU] & Index build (s) [CPU] & CSR build (s) [CPU] & Clustering (s) [CPU] & Other/I/O (s) [CPU] & Add/Build (s) [CPU] & Total (s) \\
\midrule
\texttt{BigANN-100M} & 1{,}197.56 & 49.96 & 40.78 & 43.00 & 53.26 & 187.00 & 1{,}384.56 (0.39 h) \\
\texttt{Deep1B-100M} & 1{,}176.27 & 41.85 & 64.44 & 69.00 & 53.71 & 229.00 & 1{,}405.27 (0.39 h) \\
\texttt{BigANN-1B}   & 1{,}183.46 & 529.29 & 420.50 & 380.00 & 442.21 & 1{,}772.00 & 2{,}955.47 (0.82 h) \\
\texttt{Deep1B-1B}   & 1{,}124.43 & 368.00 & 903.36 & 614.00 & 326.64 & 2{,}212.00 & 3{,}336.43 (0.93 h) \\
\texttt{GLDv2}       & 2{,}471.15 & 84.78 & 51.27 & 44.00 & 56.96 & 237.00 & 2{,}708.15 (0.75 h) \\
\bottomrule
\end{tabular}
}
\end{table}

\subsubsection{Ablations and Sensitivity}
\label{sec:ablation_sensitivity}

\paragraph{Centroid-stage early stopping.}
We ablate heap-quality-driven early stopping at the centroid stage (Step~5) on BigANN-1B with $n_q{=}10{,}000$ queries.
All retrieval hyperparameters are kept fixed ($L_{\text{sel}}{=}16$, $L_{\text{dim}}{=}22$, $E_{\text{dim}}{=}128$, and $K_c{=}3000$).
We compare early stopping enabled vs.\ disabled under two operating points (A/B) and report Recall@10, throughput (QPS),
and the refinement workload measured by candidates routed to exact refinement per query (\texttt{Cand./q}, in thousands).
Results are reported in Table~\ref{tab:ablation_early_stop}.

\paragraph{Sensitivity of $\tau_{\text{emb}}$ and $L_{\text{sim}}$ masking.}
We study sensitivity to the similarity-selection threshold $\tau_{\text{emb}}$ (selected-pair percentage) and to the pair-selection mask used in $L_{\text{sim}}$ during training.

On Deep1B-100M, we compute Pearson and Spearman correlations over $\sim\!2$M within-batch sampled pairs \emph{per setting},
using the same sampling rule and pair budget for each $\tau_{\text{emb}}\in[0.1\%,0.5\%]$.
Pairs are sampled either by thresholding in the original embedding space (\emph{Base-sampled}) or in the encoded space (\emph{Encoded-sampled});
the two procedures are applied independently and therefore may yield different pair sets.
We report correlations for the default masked objective and for the unmasked objective.
Results are reported in Table~\ref{tab:tau_mask_ablation}.

\paragraph{Bucket-local clustering.}
We ablate bucket-local clustering while keeping all other components unchanged.
When clustering is disabled, each bucket is treated as a single group (no bucket-local centroids), removing the centroid stage and routing candidates directly to refinement.
We evaluate this ablation on 1B-scale datasets (BigANN-1B and Deep1B-1B) with $n_q{=}10{,}000$ queries, and report Recall@10, QPS,
and candidates routed to exact refinement per query (\texttt{Cand./q}, in thousands), under the same retrieval configuration.
Results are reported in Table~\ref{tab:ablation_bucket_clustering}.

\section{Results and Discussion} \label{sec:results_discussion}

This section evaluates \textit{NeuRoute} in terms of retrieval quality (Recall@10), throughput (QPS), and search effort (average candidates per query), together with end-to-end index build cost (training + construction), under Euclidean distance metrics.

\textbf{Key takeaways.}
(i) \textbf{Sub-hour end-to-end build at 1B:} NeuRoute builds in 0.82\,h on BigANN-1B and 0.93\,h on Deep1B-1B, while DiskANN requires 17.96\,h and 36.47\,h, respectively.
(ii) \textbf{Competitive serving point at $\sim$90\% Recall@10:} on BigANN-1B, NeuRoute reaches 90.3\% Recall@10 at 2,414 QPS, comparable to DiskANN (90.4\% at 2,744 QPS).
(iii) \textbf{Robust across dimensions:} on GLDv2 (1536D), NeuRoute still builds in 0.75\,h under the same 2M training budget.
(iv) \textbf{Pure hashing is not a competitive ANN index at 22 bits:} under Hamming-only probing, hashing baselines either achieve low Recall@10 or become collision-dominated (e.g., TBH on Deep1B), whereas NeuRoute maintains high bucket utilization with small hotspots.

\subsection{Index Building}\label{sec:build-time}

\begin{table}[t]
\caption{
End-to-end build cost breakdown (Train $+$ Add/Build), in hours (h).
Serving footprint (GB, bytes/$10^9$) counts the on-disk serving artifacts (index metadata plus any vector payload stored by the method); the original dataset files are not counted.
Serving footprint is reported as \emph{routing-only} / \emph{routing+co-located refinement vectors}.
Peak RSS (GB) is the maximum resident memory during build.
}
\centering
\small
\setlength{\tabcolsep}{3pt}
\renewcommand{\arraystretch}{1.05}
\resizebox{\linewidth}{!}{%
\begin{tabular}{l l r r r r r}
\toprule
Dataset & Method & Train (h) & Add/Build (h) & Total (h) & Serving footprint (GB) & Peak RSS (GB) \\
\midrule
\texttt{GLDv2} & IVF-PQ   & 7.15 & 3.64 & 10.80 & 1.28  & 18.83 \\
\texttt{GLDv2} & IVF-Flat & 9.32 & 4.32 & 13.63 & 58.07 & 149.21 \\
\texttt{GLDv2} & DiskANN  & 0.00 & 1.89 & 1.89  & 76.93 & 107.92 \\
\texttt{GLDv2} & HNSW     & 0.00 & 1.79 & 1.79  & 56.11 & 100.84 \\
\texttt{GLDv2} & \textbf{NeuRoute} & \textbf{0.69} & \textbf{0.07} & \textbf{0.75} & \textbf{\mbox{0.24 / 53.98}} & \textbf{59.38} \\
\midrule
\texttt{BigANN-100M} & IVF-PQ   & 1.17 & 5.96 & 7.12 & 4.03  & 20.54 \\
\texttt{BigANN-100M} & IVF-Flat & 1.26 & 6.66 & 7.92 & 52.03 & 74.13 \\
\texttt{BigANN-100M} & DiskANN  & 0.00 & 3.02 & 3.02 & 58.51 & 79.02 \\
\texttt{BigANN-100M} & HNSW     & 0.00 & 5.94 & 5.94 & 73.03 & 107.13 \\
\texttt{BigANN-100M} & \textbf{NeuRoute} & \textbf{0.33} & \textbf{0.05} & \textbf{0.39} & \textbf{\mbox{0.78 / 12.70}} & \textbf{13.97} \\
\midrule
\texttt{BigANN-1B} & IVF-PQ & 0.63 & 31.45 & 32.08 & 40.03 & 86.69 \\
\texttt{BigANN-1B} & DiskANN & 0.00 & 17.96 & 17.96 & 341.33 & 98.32 \\
\texttt{BigANN-1B} & OPQ+IVF-PQ & 3.16 & 7.55 & 10.72 & \mbox{40.03 / 477.00} & 569.43 \\
\texttt{BigANN-1B} & \textbf{NeuRoute} & \textbf{0.33} & \textbf{0.49} & \textbf{0.82} & \textbf{\mbox{7.68 / 126.88}} & \textbf{139.57} \\
\midrule
\texttt{Deep1B-100M} & IVF-PQ   & 1.02 & 5.21 & 6.23 & 3.22  & 31.16 \\
\texttt{Deep1B-100M} & IVF-Flat & 0.63 & 3.06 & 3.69 & 39.22 & 92.74 \\
\texttt{Deep1B-100M} & DiskANN  & 0.00 & 3.98 & 3.98 & 81.92 & 104.65 \\
\texttt{Deep1B-100M} & HNSW     & 0.00 & 6.76 & 6.76 & 61.11 & 100.82 \\
\texttt{Deep1B-100M} & \textbf{NeuRoute} & \textbf{0.33} & \textbf{0.06} & \textbf{0.39} & \textbf{\mbox{0.78 / 36.54}} & \textbf{40.19} \\
\midrule
\texttt{Deep1B-1B} & IVF-PQ & 0.58 & 31.44 & 32.02 & 29.82 & 74.82 \\
\texttt{Deep1B-1B} & DiskANN & 0.00 & 36.47 & 36.47 & 819.20 & 425.26 \\
\texttt{Deep1B-1B} & OPQ+IVF-PQ & 3.71 & 6.79 & 10.51 & \mbox{24.03 / 358.00} & 419.32 \\
\texttt{Deep1B-1B} & \textbf{NeuRoute} & \textbf{0.31} & \textbf{0.61} & \textbf{0.93} & \textbf{\mbox{7.67 / 365.30}} & \textbf{386.83} \\
\bottomrule
\end{tabular}
} 
\label{tab:build_cost_breakdown}
\end{table}

Table~\ref{tab:build_cost_breakdown} reports end-to-end build cost (Train $+$ Add/Build).
For each dataset, \emph{Add/Build} starts from pre-downloaded base vectors available on the storage system and measures the wall-clock time to encode the full base set and construct the index artifacts.
NeuRoute uses a fixed 2M-vector training subset for all datasets; for GLDv2, we match the same 2M training budget across all trainable methods.
NeuRoute's trained encoder is lightweight: the saved model checkpoint is 40\,KB--4\,MB on disk, depending on the chosen encoder capacity (width/depth), latent dimension, and output code configuration.
NeuRoute \emph{Train} corresponds to encoder training on a single A100 GPU, whereas NeuRoute \emph{Add/Build} is CPU-only.
Unless otherwise stated, all CPU-side stages (including NeuRoute Add/Build and all CPU baselines) use 24 threads.
For DiskANN, index build and search use local NVMe storage as required by the method.
Dataset download and external preprocessing are excluded from timing.

\paragraph{Footprint accounting.}
We report the on-disk \emph{serving footprint}, which consists of (i) index metadata used for routing and (ii) a vector payload used for distance computation (full vectors or compressed codes); the original dataset files are not counted.
Most baselines package metadata and payload together in their default index artifacts (e.g., IVF--Flat/HNSW store full vectors; IVF--PQ/OPQ store PQ codes; DiskANN stores a graph together with a compressed vector representation).
NeuRoute separates these components for transparency: the routing index is lightweight, while exact refinement requires access to the base vectors. We therefore report routing-only / routing+co-located refinement vectors in Table~\ref{tab:build_cost_breakdown}; the latter should be compared against baseline footprints under the same “metadata+payload” definition.
Unless otherwise stated, the NeuRoute query-time results in \S6 use the co-located refinement vectors for stable throughput; omitting them reads vectors from the dataset/external vector store.

\paragraph{Build-time decomposition.}
As shown in Table~\ref{tab:NeuRoute_time_breakdown}, NeuRoute Add/Build decomposes into index build (full-base encoding plus a lightweight calibration/evaluation step), CSR file construction, bucket-local clustering, and residual I/O.
Across datasets, refinement-cache construction and bucket-local clustering account for a substantial fraction of Add/Build (roughly 40--69\%), while the calibration/evaluation portion of index build is lightweight relative to the other components.

\paragraph{Build-time advantage.}
Across datasets, NeuRoute substantially reduces end-to-end build time under matched training budgets.
On GLDv2, NeuRoute builds in 0.75\,h versus 10.80\,h for IVF--PQ ($\approx$14--15$\times$ faster).
On 100M-scale benchmarks, NeuRoute builds in 0.39\,h versus 6.23--7.12\,h for IVF--PQ ($\approx$16--18$\times$ faster).
Most notably, at 1B scale NeuRoute completes end-to-end build in 0.82\,h on BigANN-1B and 0.93\,h on Deep1B-1B. This is $\approx$22$\times$ and $\approx$39$\times$ faster than DiskANN, respectively, and $\approx$11--13$\times$ faster than OPQ+IVF--PQ (Train+Add/Build).

\section{Retrieval Results}
\label{sec:retrieval_results}

\begin{figure}[t]
  \centering
\includegraphics[width=0.85\textwidth]   {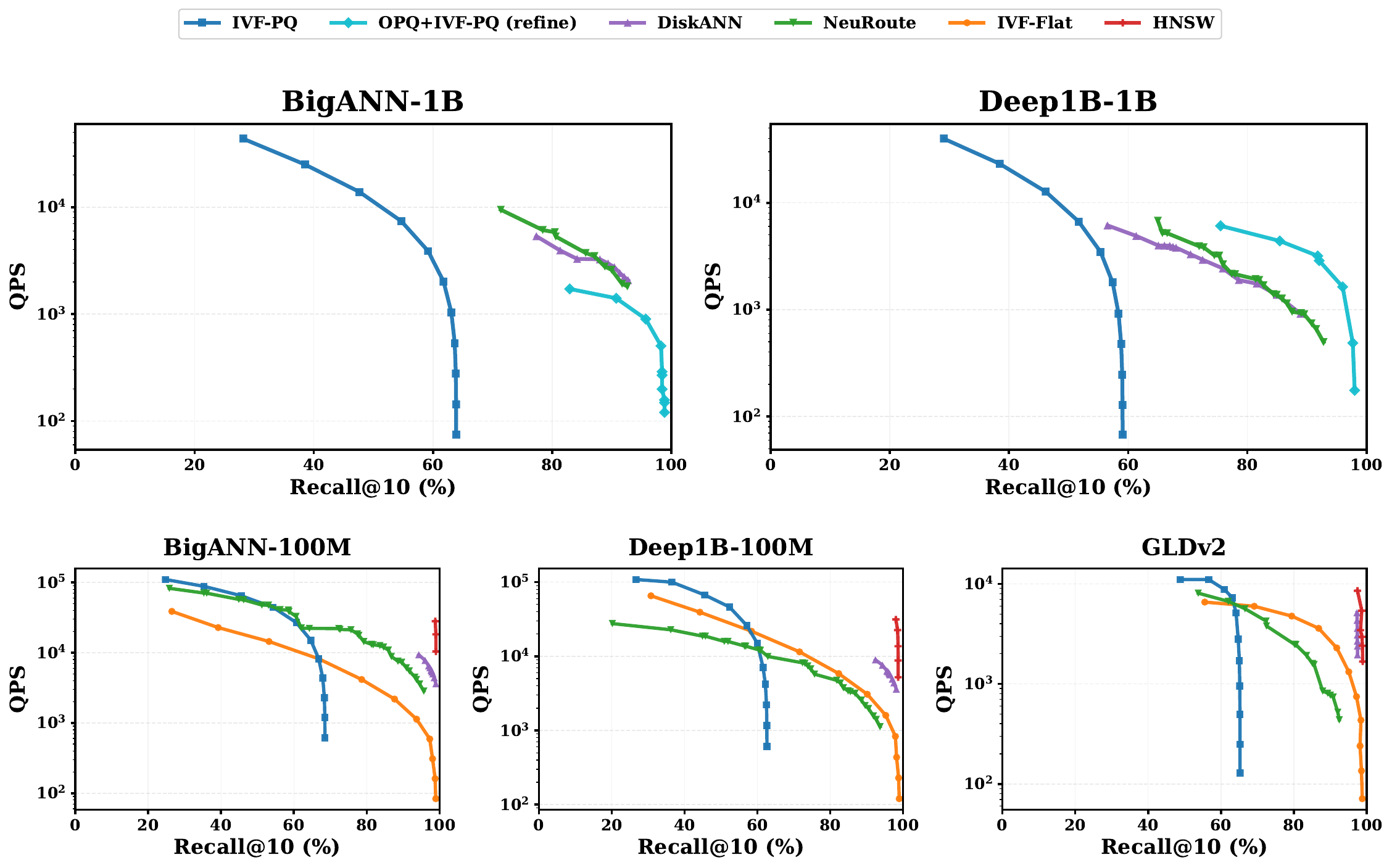}
  \caption{Recall@10--QPS trade-offs on \texttt{BigANN-1B}, \texttt{Deep1B-1B}, \texttt{BigANN-100M}, \texttt{Deep1B-100M}, and \texttt{GLDv2}. Each curve shows the Pareto frontier obtained by sweeping the main accuracy knob of each baseline (e.g., \texttt{nprobe} for IVF variants, \texttt{efSearch} for HNSW, and the search-budget parameters for DiskANN and rerank/refine variants). Higher is better.}
  \label{fig:recall_qps_tradeoff}
\end{figure}

\subsection{Baseline Accuracy--Throughput Trade-offs}
\label{subsec:baseline_tradeoffs}

Figure~\ref{fig:recall_qps_tradeoff} summarizes the accuracy--throughput envelopes of representative ANN indexes.
Across datasets, IVF-based methods follow the standard pattern: very high throughput at low recall when probing few lists, followed by steep QPS degradation as recall increases.
For example, \texttt{IVF-PQ} is extremely fast in the low-recall regime (e.g., $110{,}360$ QPS at $24.7\%$ recall on \texttt{BigANN-100M} and $43{,}534$ QPS at $28.2\%$ recall on \texttt{BigANN-1B}), but its maximum recall remains limited under a fixed code budget (e.g., $68.5\%$ on \texttt{BigANN-100M} and $63.9\%$ on \texttt{BigANN-1B}).
In contrast, \texttt{IVF-Flat} can approach near-exact recall, but only by scanning substantially more candidates, which reduces throughput (e.g., $\approx 309$ QPS at $99.1\%$ recall on \texttt{BigANN-100M}, and $\approx 240$ QPS at $99.2\%$ recall on \texttt{GLDv2}).

Graph-based indexes dominate the near-saturated recall regime on the 100M-scale settings.
On \texttt{BigANN-100M}, HNSW reaches near-perfect recall with strong throughput (e.g., $28{,}084$ QPS at $99.8\%$ recall), while DiskANN offers a different trade-off (e.g., $\approx 4{,}996$ QPS at $\approx 98.2\%$ recall).
On \texttt{Deep1B-100M}, HNSW similarly reaches high recall with high throughput (e.g., $31{,}361$ QPS at $99.0\%$ recall), and DiskANN achieves $\approx 3{,}617$ QPS at $\approx 98.1\%$ recall.
On \texttt{GLDv2}, DiskANN achieves high recall with strong throughput (e.g., $5{,}152$ QPS at $99.4\%$ recall), while HNSW can be tuned either for higher QPS at slightly lower recall (e.g., $8{,}534$ QPS at $97.5\%$) or for near-perfect recall at lower throughput (e.g., $1{,}673$ QPS at $99.9\%$).

\subsection{Smaller-Scale and High-Dimensional Results}
\label{subsec:small_and_highdim}

While graph-based methods set a strong upper bound on moderate-scale benchmarks, NeuRoute targets a complementary operating point: \emph{mid-to-high recall with predictable, budgeted candidate generation and substantially lower end-to-end build cost} (Table~\ref{tab:build_cost_breakdown}).

\paragraph{100M-scale results.}
On \texttt{BigANN-100M}, NeuRoute reaches $88.6\%$ Recall@10 at $7{,}565$ QPS and can extend to $95.9\%$ Recall@10 at $2{,}665$ QPS under the same sweep, while building end-to-end in $0.39$ hours (Train+Add/Build).
On \texttt{Deep1B-100M}, NeuRoute reaches $88.6\%$ Recall@10 at $2{,}559$ QPS and extends to $94.0\%$ Recall@10 at $1{,}054$ QPS, also with a $0.39$ hour end-to-end build.
These results show that NeuRoute remains competitive in the mid-to-high recall regime on 100M-scale datasets while retaining a much lower build-time footprint than graph/quantization pipelines.

\paragraph{High-dimensional \texttt{GLDv2} (1536D)}
On \texttt{GLDv2}, NeuRoute remains usable at high dimensionality: it reaches $72.4\%$ Recall@10 at $4{,}251$ QPS, and reaches $88.1\%$ Recall@10 at $814$ QPS; pushing recall further to $92.7\%$ yields $425$ QPS.
Meanwhile, NeuRoute builds end-to-end in $0.75$ hours on the same node (Table~\ref{tab:build_cost_breakdown}), demonstrating that the approach continues to function in modern high-dimensional embedding spaces, even though graph-based baselines can reach higher near-saturated recall on this dataset.

\paragraph{Pure Hashing at 22 Bits (Context)}
\label{subsec:hashing_context}

Although NeuRoute uses short binary addresses, it is \emph{not} a pure hashing index:
query processing interleaves budget-aware bucket routing and centroid-stage filtering
before exact refinement (\S\ref{subsec:retrieval}).
For context, we evaluate standard 22-bit hashing baselines (LSH, ITQ, TBH) under
\emph{Hamming-only} probing with the same bit budget ($b=22$, stored as 3 bytes/vector when bit-packed).
All methods are evaluated under the same harness (same query set, distance metric, and exact refinement implementation).
On the high-dimensional 10M-scale setting (GLDv2, 1536D), these pure-hashing baselines exhibit a weak
recall--throughput trade-off under short codes. In this regime, NeuRoute achieves consistently better
accuracy--throughput performance under the same 22-bit budget. Table~\ref{tab:gldv2_hash_vs_NeuRoute_main}
reports a compact comparison.

For fairness in this subsection, NeuRoute is evaluated using \emph{only} the learned 22-bit codes
(with bucket-local clustering disabled) to isolate the effect of short-code routing under the same bit budget.
We omit version-specific hashing frontiers and bucket-structure diagnostics to prioritize the 1B-scale comparisons against strong ANN baselines (DiskANN and quantization-based methods).

\begin{table}[t]
\centering
\caption{GLDv2 (1536D) comparison under the same evaluation harness ($n_q{=}10{,}000$).
For fairness, NeuRoute is evaluated using only the learned 22-bit codes (no bucket-local clustering).}
\footnotesize
\setlength{\tabcolsep}{7pt}
\renewcommand{\arraystretch}{1.05}
\begin{tabular}{lrr}
\toprule
Method (22b) & Recall@10 (\%) & QPS \\
\midrule
LSH & 37.55 & 82.94 \\
ITQ & 57.27 & 111.17 \\
TBH & 74.99 & 18.52 \\
\midrule
NeuRoute (no clustering) & 93.86 & 111.40 \\
\bottomrule
\end{tabular}

\label{tab:gldv2_hash_vs_NeuRoute_main}
\end{table}

\paragraph{GPU-only reference (CAGRA)}
We additionally ran the GPU graph baseline CAGRA~\cite{ootomo2024cagrahighlyparallelgraph} as a throughput reference.
In our environment, CAGRA is device-memory limited (A100 40GB) and can be built only up to 80M vectors, so we do not include it in the main 100M/1B Recall@10--QPS comparisons; the 80M results are reported in Appendix~\ref{app:cagra_reference}.

\subsection{Full Billion-Scale Results}
\label{subsec:full_1b_results}

The billion-scale datasets highlight the limitations of pure \texttt{IVF-PQ} at fixed code size and the importance of better candidate generation.
On \texttt{BigANN-1B}, \texttt{IVF-PQ} tops out at $63.9\%$ recall in our sweep, while DiskANN, NeuRoute, and reranked OPQ+IVF-PQ cover the high-recall regime.
At an operating point near $90\%$ Recall@10, NeuRoute sustains $2{,}414$ QPS at $90.3\%$ recall, comparable to DiskANN ($2{,}744$ QPS at $90.4\%$ recall) and substantially faster than OPQ+IVF-PQ (refine) ($1{,}406$ QPS at $90.8\%$ recall).
Crucially, this operating point is achieved with dramatically lower build cost: NeuRoute builds in $0.82$ hours versus $17.96$ hours for DiskANN and $10.72$ hours for OPQ+IVF-PQ (Train+Add/Build).
DiskANN and NeuRoute reach similar ceilings in the low-90s recall range (DiskANN: $92.7\%$ at $2{,}071$ QPS; NeuRoute: $93.0\%$ at $1{,}707$ QPS), whereas OPQ+IVF-PQ (refine) is the only baseline here that pushes to $\approx 99.9\%$ recall, at the cost of much lower throughput ($\approx 121$ QPS at $99.9\%$ recall).

On \texttt{Deep1B-1B}, OPQ+IVF-PQ (refine) provides the strongest high-recall frontier in this set of runs, reaching $3{,}342$ QPS at $91.2\%$ recall and extending to $99.4\%$ recall at $248$ QPS.
DiskANN reaches $88.9\%$ Recall@10 at $915$ QPS, while NeuRoute reaches $90.1\%$ Recall@10 at $842$ QPS; critically, NeuRoute builds in $0.93$ hours versus DiskANN's $36.47$ hours.
Taken together, the two 1B datasets indicate that NeuRoute consistently reaches the low-90s recall regime with competitive throughput, while delivering \emph{sub-hour} end-to-end build time at billion scale; reranked OPQ+IVF-PQ can extend to near-perfect recall when the additional refinement cost is acceptable.

\begin{table}[t]
\centering
\small
\setlength{\tabcolsep}{4pt}
\renewcommand{\arraystretch}{1.05}
\caption{Billion-scale operating points \emph{around} 90\% Recall@10 (single-node).
For each method, we report the highest QPS among configurations whose Recall@10 falls in a narrow band around 0.90
(here: $[0.88, 0.92]$). Build time is end-to-end Train+Add/Build from Table~\ref{tab:build_cost_breakdown}.}
\label{tab:1b_operating_points_around90}
\begin{tabular}{l l r r r}
\toprule
Dataset & Method & Recall@10 & QPS & Build time (h) \\
\midrule
\texttt{BigANN-1B} & DiskANN & 0.9040 & 2,744 & 17.96 \\
\texttt{BigANN-1B} & OPQ+IVF-PQ (refine) & 0.9080 & 1,406 & 10.72 \\
\texttt{BigANN-1B} & \textbf{NeuRoute} & 0.9030 & \textbf{2,414} & \textbf{0.82} \\
\midrule
\texttt{Deep1B-1B} & DiskANN & 0.8891 & 914.68 & 36.47 \\
\texttt{Deep1B-1B} & OPQ+IVF-PQ (refine) & 0.9120 & 3,342 & 10.51 \\
\texttt{Deep1B-1B} & \textbf{NeuRoute} & 0.9010 & \textbf{842} & \textbf{0.93} \\
\bottomrule
\end{tabular}
\end{table}

\subsection{Ablation Studies}
We ablate three components of \textit{NeuRoute}: heap-quality-driven early stopping at query time,
the similarity-selection threshold $\tau_{\text{emb}}$ and the pair-selection mask in $L_{\text{sim}}$ during training,
and bucket-local clustering in the retrieval pipeline.

\paragraph{Heap-quality-driven early stopping.}
We ablate heap-quality-driven early stopping on BigANN-1B with $n_q=10{,}000$ queries.
All retrieval hyperparameters are kept fixed: $L_{\text{sel}}=16$, $L_{\text{dim}}=22$, $E_{\text{dim}}=128$, and $K_c=3000$.
We report throughput (QPS), Recall@10, and the typical refinement workload in terms of candidates routed to exact refinement
(\texttt{Cand./q} in thousands; computed from the query-level refinement statistics).
Table~\ref{tab:ablation_early_stop} compares two operating points and shows a consistent trend:
early stopping improves QPS by $1.5\times$--$1.7\times$ while reducing refined candidates by $\sim$16--19\%,
with a negligible absolute Recall@10 drop ($<10^{-3}$).

\begin{table}[t]
\centering
\small
\setlength{\tabcolsep}{3pt}
\renewcommand{\arraystretch}{1.05}
\caption{Ablation of heap-quality-driven early stopping on BigANN-1B ($n_q=10{,}000$).}
\label{tab:ablation_early_stop}
\begin{tabular}{llrrr}
\toprule
Op. & Variant & Recall@10 & QPS & Cand./q (k) \\
\midrule
A & No early stop & 0.90335 & 1440 & 249 \\
A & Early stop    & 0.90260 & 2414 & 201 \\
\midrule
B & No early stop & 0.90038 & 1682 & 230 \\
B & Early stop    & 0.89965 & 2595 & 194 \\
\bottomrule
\end{tabular}
\end{table}

\paragraph{Sensitivity to $\tau_{\text{emb}}$ and the pair-selection mask in $L_{\text{sim}}$.}
We study sensitivity to the top-quantile threshold $\tau_{\text{emb}}$ (fraction of selected within-batch pairs) and the pair-selection mask used in $L_{\text{sim}}$.
We compute Pearson/Spearman correlations over $2{,}096{,}650$ within-batch near-neighbor pairs, sampled separately in the original embedding space (\emph{Base-sampled}) or in the encoded space (\emph{Encoded-sampled}) using the same top-quantile rule (thus not necessarily the same pairs).
As shown in Table~\ref{tab:tau_mask_ablation}, correlations remain stable when sweeping $\tau_{\text{emb}}\in[0.1\%,0.5\%]$ under the default masked objective.
In contrast, removing the latent-space mask (run-1) or removing masking entirely (run-2) substantially degrades correlations for encoded-sampled pairs, indicating that masking is important for filtering noisy pairs and preserving local neighborhood structure during training.

\begin{table}[t]
\centering
\small
\setlength{\tabcolsep}{4pt}
\renewcommand{\arraystretch}{1.05}
\caption{Sensitivity to the training threshold $\tau_{\text{emb}}$ and the pair-selection mask on Deep1B-100M (database-sampled). Pearson/Spearman correlations are evaluated on a fixed set of $2{,}096{,}650$ within-batch near-neighbor pairs, sampled as the top 0.5\% most similar pairs separately in the base and encoded spaces (thus not necessarily the same pairs).}

\label{tab:tau_mask_ablation}
\begin{tabular}{lrrrr}
\toprule
\multirow{2}{*}{$\tau_{\text{emb}}$ (\%)} & \multicolumn{2}{c}{Base-sampled pairs} & \multicolumn{2}{c}{Encoded-sampled pairs} \\
\cmidrule(lr){2-3}\cmidrule(lr){4-5}
 & Pearson & Spearman & Pearson & Spearman \\
\midrule
0.1 & 0.8335 & 0.7722 & 0.8027 & 0.7362 \\
0.2 & 0.8422 & 0.7789 & 0.8162 & 0.7429 \\
0.3 & 0.8458 & 0.7813 & 0.8190 & 0.7457 \\
0.5 & 0.8460 & 0.7800 & 0.8130 & 0.7430 \\
\midrule
0.5 (w/o mask, run-1) & 0.8368 & 0.7535 & 0.5496 & 0.5617 \\
N/A (w/o mask, run-2) & 0.7391 & 0.6513 & 0.6701 & 0.5774 \\
\bottomrule
\end{tabular}
\end{table}

\paragraph{Bucket-local clustering.}
We ablate bucket-local clustering while keeping all other components unchanged.
When clustering is disabled, each bucket is treated as a single group (no bucket-local centroids),
so the centroid-stage is removed and candidates are routed directly to refinement.
Table~\ref{tab:ablation_bucket_clustering} shows that bucket-local clustering consistently reduces refinement load,
leading to substantial throughput gains.
On BigANN-1B, clustering improves QPS by $4.44\times$ (2414 vs.\ 543.6),
increases Recall@10 by $+1.48$ points (90.26\% vs.\ 88.78\%),
and reduces candidates/query by $4.68\times$ (201k vs.\ 940k).
On Deep1B-1B, clustering improves QPS by $4.57\times$ (756.9 vs.\ 165.5) and reduces candidates/query by $3.86\times$
(351k vs.\ 1353k) with a marginal Recall@10 gain of $+0.14$ points (90.14\% vs.\ 90.01\%).
Unless otherwise stated, we enable bucket-local clustering; we disable it in the 22-bit hashing comparison to isolate
the effect of short-code routing under the same bit budget.

\begin{table}[t]
\centering
\small
\setlength{\tabcolsep}{3pt}
\renewcommand{\arraystretch}{1.05}
\caption{Ablation of bucket-local clustering on 1B-scale datasets ($n_q{=}10{,}000$ queries).
QPS is measured under the same retrieval configuration.
Cand./q (k) is the number of vectors routed to exact refinement per query, reported in thousands.}
\label{tab:ablation_bucket_clustering}
\begin{tabular}{llrrr}
\toprule
Dataset & Variant & Recall@10 & QPS & Cand./q (k) \\
\midrule
\texttt{BigANN-1B} & w/ clustering  & 0.90260 & 2414  & 201 \\
\texttt{BigANN-1B} & no clustering  & 0.88780 & 543.6 & 940 \\
\midrule
\texttt{Deep1B-1B} & w/ clustering  & 0.90144 & 756.9 & 351 \\
\texttt{Deep1B-1B} & no clustering  & 0.90007 & 165.5 & 1353 \\
\bottomrule
\end{tabular}
\end{table}

\paragraph{Takeaways.}
(1) Centroid-stage early stopping is a safe efficiency knob: it yields a consistent $1.5\times$--$1.7\times$ QPS improvement by reducing refinement workload ($\sim$16--19\% fewer candidates) with a negligible Recall@10 drop ($<10^{-3}$).
(2) The $L_{\text{sim}}$ training signal is robust to $\tau_{\text{emb}}$: correlations remain stable when sweeping $\tau_{\text{emb}}\in[0.1\%,0.5\%]$ under the masked objective. Masking is essential, especially for encoded-space sampling: removing the latent-space mask causes a large correlation collapse (e.g., Encoded-sampled Pearson/Spearman $\approx 0.81/0.74 \rightarrow 0.55/0.56$).
(3) Bucket-local clustering is critical for throughput at billion scale: it cuts refinement load by $3.9\times$--$4.7\times$ and improves QPS by $4.4\times$--$4.6\times$, while preserving (and sometimes improving) Recall@10 (e.g., +1.48 points on BigANN-1B and +0.14 on Deep1B-1B). Overall, NeuRoute's gains primarily come from reducing refinement work through centroid-stage structure, calibrated gating, and early termination.

\section{Conclusion}
\label{sec:conclusion}

This paper presented NeuRoute, an encoder-based hashing framework for scalable $k$-nearest-neighbor search in large vector databases. NeuRoute learns compact binary codes for fast, memory-efficient candidate generation. It combines logit-guided routing with calibrated gating and heap-quality-driven early stopping to achieve strong accuracy--throughput trade-offs at billion scale. In our experiments, NeuRoute attains 90\%+ Recall@10 on 1B-vector benchmarks while enabling sub-hour end-to-end builds on a single machine. Encoder training is GPU-accelerated, whereas index construction and online retrieval are CPU-based.

NeuRoute also admits orthogonal system extensions, such as PQ-based reranking in the refinement stage, which may further improve the tradeoff curve.
Future work will study how to support additional quantization variants (e.g., OPQ+PQ) within the same refinement interface.
We also plan to harden NeuRoute for production deployment. The current implementation and evaluation artifacts are publicly available to support reproducibility.

\section*{Acknowledgment}
This research is supported by the Arkansas High Performance Computing Center, which is funded through multiple National Science Foundation grants and the Arkansas Economic Development Commission. This project was also partially supported by the National Science Foundation under Award No. OIA-1946391.

\bibliographystyle{unsrtnat}
\bibliography{references}

\begin{thebibliography}{49}
\providecommand{\natexlab}[1]{#1}
\providecommand{\url}[1]{\texttt{#1}}
\expandafter\ifx\csname urlstyle\endcsname\relax
  \providecommand{\doi}[1]{doi: #1}\else
  \providecommand{\doi}{doi: \begingroup \urlstyle{rm}\Url}\fi

\bibitem[Taipalus(2024)]{ref15}
T.~Taipalus.
\newblock Vector database management systems: Fundamental concepts, use-cases,
  and current challenges.
\newblock \emph{Cognitive Systems Research}, 85:\penalty0 Article 101216, 2024.
\newblock URL
  \url{https://www.sciencedirect.com/science/article/pii/S1389041724000093}.

\bibitem[Wang et~al.(2021{\natexlab{a}})Wang, Yi, Guo, Jin, Xu, Li, Wang, Guo,
  Li, Xu, Yu, Yuan, Zou, Long, Cai, Li, Zhang, Mo, Gu, Jiang, Wei, and
  Xie]{10.1145/3448016.3457550}
Jianguo Wang, Xiaomeng Yi, Rentong Guo, Hai Jin, Peng Xu, Shengjun Li, Xiangyu
  Wang, Xiangzhou Guo, Chengming Li, Xiaohai Xu, Kun Yu, Yuxing Yuan, Yinghao
  Zou, Jiquan Long, Yudong Cai, Zhenxiang Li, Zhifeng Zhang, Yihua Mo, Jun Gu,
  Ruiyi Jiang, Yi~Wei, and Charles Xie.
\newblock Milvus: A purpose-built vector data management system.
\newblock In \emph{Proceedings of the 2021 International Conference on
  Management of Data}, SIGMOD '21, page 2614–2627, New York, NY, USA,
  2021{\natexlab{a}}. Association for Computing Machinery.
\newblock ISBN 9781450383431.
\newblock \doi{10.1145/3448016.3457550}.
\newblock URL \url{https://doi.org/10.1145/3448016.3457550}.

\bibitem[Simhadri et~al.(2022)Simhadri, Williams, Aum{\"u}ller, Douze, Babenko,
  Baranchuk, Chen, Hosseini, Krishnaswamy, Srinivasa, Subramanya, and
  Wang]{simhadri2022bigann}
Harsha~Vardhan Simhadri, George Williams, Martin Aum{\"u}ller, Matthijs Douze,
  Artem Babenko, Dmitry Baranchuk, Qi~Chen, Lucas Hosseini, Ravishankar
  Krishnaswamy, Gopal Srinivasa, Suhas~Jayaram Subramanya, and Jingdong Wang.
\newblock Results of the neurips’21 challenge on billion-scale approximate
  nearest neighbor search.
\newblock In \emph{Proceedings of the NeurIPS 2021 Competitions and
  Demonstrations Track}, volume 176 of \emph{PMLR}, pages 177--189, 2022.
\newblock URL \url{https://proceedings.mlr.press/v176/simhadri22a.html}.

\bibitem[Malkov and Yashunin(2020)]{33}
Yu~A. Malkov and Dmitry~A. Yashunin.
\newblock Efficient and robust approximate nearest neighbor search using
  hierarchical navigable small world graphs.
\newblock \emph{IEEE Transactions on Pattern Analysis and Machine
  Intelligence}, 42\penalty0 (4):\penalty0 824--836, 2020.
\newblock \doi{10.1109/TPAMI.2018.2889473}.

\bibitem[Jayaram~Subramanya et~al.(2019)Jayaram~Subramanya, Devvrit, Simhadri,
  Krishnawamy, and Kadekodi]{NEURIPS2019_09853c7f}
Suhas Jayaram~Subramanya, Fnu Devvrit, Harsha~Vardhan Simhadri, Ravishankar
  Krishnawamy, and Rohan Kadekodi.
\newblock {DiskANN}: Fast accurate billion-point nearest neighbor search on a
  single node.
\newblock In H.~Wallach, H.~Larochelle, A.~Beygelzimer, F.~d\textquotesingle
  Alch\'{e}-Buc, E.~Fox, and R.~Garnett, editors, \emph{Advances in Neural
  Information Processing Systems}, volume~32. Curran Associates, Inc., 2019.
\newblock URL
  \url{https://proceedings.neurips.cc/paper_files/paper/2019/file/09853c7fb1d3f8ee67a61b6bf4a7f8e6-Paper.pdf}.

\bibitem[Wang et~al.(2021{\natexlab{b}})Wang, Xu, Yue, and
  Wang]{wang2021comprehensivesurveyexperimentalcomparison}
Mengzhao Wang, Xiaoliang Xu, Qiang Yue, and Yuxiang Wang.
\newblock A comprehensive survey and experimental comparison of graph-based
  approximate nearest neighbor search, 2021{\natexlab{b}}.
\newblock URL \url{https://arxiv.org/abs/2101.12631}.

\bibitem[Chen et~al.(2021)Chen, Zhao, Wang, Li, Liu, Li, Yang, and
  Wang]{NEURIPS2021_299dc35e}
Qi~Chen, Bing Zhao, Haidong Wang, Mingqin Li, Chuanjie Liu, Zengzhong Li, Mao
  Yang, and Jingdong Wang.
\newblock {SPANN}: Highly-efficient billion-scale approximate nearest
  neighborhood search.
\newblock In M.~Ranzato, A.~Beygelzimer, Y.~Dauphin, P.S. Liang, and J.~Wortman
  Vaughan, editors, \emph{Advances in Neural Information Processing Systems},
  volume~34, pages 5199--5212. Curran Associates, Inc., 2021.
\newblock URL
  \url{https://proceedings.neurips.cc/paper_files/paper/2021/file/299dc35e747eb77177d9cea10a802da2-Paper.pdf}.

\bibitem[Zhao et~al.(2023)Zhao, Tian, Huang, Zheng, and
  Zhou]{10.14778/3594512.3594527}
Xi~Zhao, Yao Tian, Kai Huang, Bolong Zheng, and Xiaofang Zhou.
\newblock Towards efficient index construction and approximate nearest neighbor
  search in high-dimensional spaces.
\newblock \emph{Proc. VLDB Endow.}, 16\penalty0 (8):\penalty0 1979–1991,
  April 2023.
\newblock ISSN 2150-8097.
\newblock \doi{10.14778/3594512.3594527}.
\newblock URL \url{https://doi.org/10.14778/3594512.3594527}.

\bibitem[Weiss et~al.(2008)Weiss, Torralba, and Fergus]{NIPS2008_d58072be}
Yair Weiss, Antonio Torralba, and Rob Fergus.
\newblock Spectral hashing.
\newblock In D.~Koller, D.~Schuurmans, Y.~Bengio, and L.~Bottou, editors,
  \emph{Advances in Neural Information Processing Systems}, volume~21. Curran
  Associates, Inc., 2008.
\newblock URL
  \url{https://proceedings.neurips.cc/paper_files/paper/2008/file/d58072be2820e8682c0a27c0518e805e-Paper.pdf}.

\bibitem[Gong and Lazebnik(2011)]{inproceedings2011}
Yunchao Gong and Svetlana Lazebnik.
\newblock Iterative quantization: A procrustean approach to learning binary
  codes.
\newblock In \emph{2011 IEEE Conference on Computer Vision and Pattern
  Recognition (CVPR)}, pages 817--824, 06 2011.
\newblock \doi{10.1109/CVPR.2011.5995432}.

\bibitem[Erin~Liong et~al.(2015)Erin~Liong, Lu, Wang, Moulin, and
  Zhou]{Liong_2015_CVPR}
Venice Erin~Liong, Jiwen Lu, Gang Wang, Pierre Moulin, and Jie Zhou.
\newblock Deep hashing for compact binary codes learning.
\newblock In \emph{Proceedings of the IEEE Conference on Computer Vision and
  Pattern Recognition (CVPR)}, June 2015.

\bibitem[Jafari and Nagarkar(2021)]{26}
Omid Jafari and Parth Nagarkar.
\newblock Experimental analysis of locality sensitive hashing techniques for
  high-dimensional approximate nearest neighbor searches.
\newblock In Miao Qiao, Gottfried Vossen, Sen Wang, and Lei Li, editors,
  \emph{Databases Theory and Applications}, pages 62--73, Cham, 2021. Springer
  International Publishing.
\newblock ISBN 978-3-030-69377-0.

\bibitem[Norouzi et~al.(2012)Norouzi, Fleet, and
  Salakhutdinov]{NIPS2012_59b90e10}
Mohammad Norouzi, David~J Fleet, and Russ~R Salakhutdinov.
\newblock Hamming distance metric learning.
\newblock In F.~Pereira, C.J. Burges, L.~Bottou, and K.Q. Weinberger, editors,
  \emph{Advances in Neural Information Processing Systems}, volume~25. Curran
  Associates, Inc., 2012.
\newblock URL
  \url{https://proceedings.neurips.cc/paper_files/paper/2012/file/59b90e1005a220e2ebc542eb9d950b1e-Paper.pdf}.

\bibitem[Pan et~al.(2023)Pan, Wang, and
  Li]{pan2023surveyvectordatabasemanagement}
James~Jie Pan, Jianguo Wang, and Guoliang Li.
\newblock Survey of vector database management systems, 2023.
\newblock URL \url{https://arxiv.org/abs/2310.14021}.

\bibitem[Johnson et~al.(2017)Johnson, Douze, and
  Jégou]{johnson2017billionscalesimilaritysearchgpus}
Jeff Johnson, Matthijs Douze, and Hervé Jégou.
\newblock Billion-scale similarity search with {GPUs}, 2017.
\newblock URL \url{https://arxiv.org/abs/1702.08734}.

\bibitem[Douze et~al.(2024)Douze, Guzhva, Deng, Johnson, Szilvasy, Mazaré,
  Lomeli, Hosseini, and Jégou]{ref40}
Matthijs Douze, Alexandr Guzhva, Chengqi Deng, Jeff Johnson, Gergely Szilvasy,
  Pierre-Emmanuel Mazaré, Maria Lomeli, Lucas Hosseini, and Hervé Jégou.
\newblock The faiss library.
\newblock \emph{arXiv preprint arXiv:2401.08281}, 2024.
\newblock URL \url{https://arxiv.org/abs/2401.08281}.

\bibitem[Jégou et~al.(2011)Jégou, Douze, and Schmid]{5432202}
Herve Jégou, Matthijs Douze, and Cordelia Schmid.
\newblock Product quantization for nearest neighbor search.
\newblock \emph{IEEE Transactions on Pattern Analysis and Machine
  Intelligence}, 33\penalty0 (1):\penalty0 117--128, 2011.
\newblock \doi{10.1109/TPAMI.2010.57}.

\bibitem[Ge et~al.(2014)Ge, He, Ke, and Sun]{6678503}
Tiezheng Ge, Kaiming He, Qifa Ke, and Jian Sun.
\newblock Optimized product quantization.
\newblock \emph{IEEE Transactions on Pattern Analysis and Machine
  Intelligence}, 36\penalty0 (4):\penalty0 744--755, 2014.
\newblock \doi{10.1109/TPAMI.2013.240}.

\bibitem[Simhadri and Contributors(2021)]{bigann_readme}
Harsha~Vardhan Simhadri and Big ANN~Benchmarks Contributors.
\newblock Big ann benchmarks: Neurips’21 t1/t2 readme.
\newblock GitHub repository, 2021.
\newblock URL \url{https://github.com/harsha-simhadri/big-ann-benchmarks}.
\newblock Accessed: 2025-09-09.

\bibitem[Elliott and Clark(2024)]{27}
Owen~Pendrigh Elliott and Jesse Clark.
\newblock The impacts of data, ordering, and intrinsic dimensionality on recall
  in hierarchical navigable small worlds, 2024.
\newblock URL \url{https://arxiv.org/abs/2405.17813}.

\bibitem[Lu et~al.(2021)Lu, Kudo, Xiao, and Ishikawa]{34}
Kejing Lu, Mineichi Kudo, Chuan Xiao, and Yoshiharu Ishikawa.
\newblock {HVS}: Hierarchical graph structure based on voronoi diagrams for
  solving approximate nearest neighbor search.
\newblock \emph{PVLDB}, 15\penalty0 (2):\penalty0 246--258, 2021.
\newblock URL \url{http://www.vldb.org/pvldb/vol15/p246-lu.pdf}.

\bibitem[Fu et~al.(2017)Fu, Wang, and Cai]{CongFu}
Cong Fu, Changxu Wang, and Deng Cai.
\newblock Fast approximate nearest neighbor search with navigating
  spreading-out graphs.
\newblock \emph{CoRR}, abs/1707.00143, 2017.
\newblock URL \url{http://arxiv.org/abs/1707.00143}.

\bibitem[Ootomo et~al.(2024)Ootomo, Naruse, Nolet, Wang, Feher, and
  Wang]{ootomo2024cagrahighlyparallelgraph}
Hiroyuki Ootomo, Akira Naruse, Corey Nolet, Ray Wang, Tamas Feher, and Yong
  Wang.
\newblock {CAGRA}: Highly parallel graph construction and approximate nearest
  neighbor search for {GPUs}, 2024.
\newblock URL \url{https://arxiv.org/abs/2308.15136}.

\bibitem[Luo et~al.(2023)Luo, Wang, Wu, Chen, Deng, Huang, and Hua]{hash}
Xiao Luo, Haixin Wang, Daqing Wu, Chong Chen, Minghua Deng, Jianqiang Huang,
  and Xian-Sheng Hua.
\newblock A survey on deep hashing methods.
\newblock \emph{ACM Trans. Knowl. Discov. Data}, 17\penalty0 (1), February
  2023.
\newblock ISSN 1556-4681.
\newblock \doi{10.1145/3532624}.
\newblock URL \url{https://doi.org/10.1145/3532624}.

\bibitem[Lai et~al.(2015)Lai, Pan, Liu, and Yan]{Lai_2015}
Hanjiang Lai, Yan Pan, Ye~Liu, and Shuicheng Yan.
\newblock Simultaneous feature learning and hash coding with deep neural
  networks.
\newblock In \emph{2015 IEEE Conference on Computer Vision and Pattern
  Recognition (CVPR)}. IEEE, June 2015.
\newblock \doi{10.1109/cvpr.2015.7298947}.
\newblock URL \url{http://dx.doi.org/10.1109/CVPR.2015.7298947}.

\bibitem[Charikar(2002)]{10.1145/509907.509965}
Moses~S. Charikar.
\newblock Similarity estimation techniques from rounding algorithms.
\newblock In \emph{Proceedings of the Thirty-Fourth Annual ACM Symposium on
  Theory of Computing}, STOC '02, page 380–388, New York, NY, USA, 2002.
  Association for Computing Machinery.
\newblock ISBN 1581134959.
\newblock \doi{10.1145/509907.509965}.
\newblock URL \url{https://doi.org/10.1145/509907.509965}.

\bibitem[Datar et~al.(2004)Datar, Immorlica, Indyk, and
  Mirrokni]{10.1145/997817.997857}
Mayur Datar, Nicole Immorlica, Piotr Indyk, and Vahab~S. Mirrokni.
\newblock Locality-sensitive hashing scheme based on p-stable distributions.
\newblock In \emph{Proceedings of the Twentieth Annual Symposium on
  Computational Geometry}, SCG '04, page 253–262, New York, NY, USA, 2004.
  Association for Computing Machinery.
\newblock ISBN 1581138857.
\newblock \doi{10.1145/997817.997857}.
\newblock URL \url{https://doi.org/10.1145/997817.997857}.

\bibitem[Indyk and Motwani(1998)]{10.1145/276698.276876}
Piotr Indyk and Rajeev Motwani.
\newblock Approximate nearest neighbors: towards removing the curse of
  dimensionality.
\newblock In \emph{Proceedings of the Thirtieth Annual ACM Symposium on Theory
  of Computing}, STOC '98, page 604–613, New York, NY, USA, 1998. Association
  for Computing Machinery.
\newblock ISBN 0897919629.
\newblock \doi{10.1145/276698.276876}.
\newblock URL \url{https://doi.org/10.1145/276698.276876}.

\bibitem[Mau and Huynh(2021)]{31}
Toan~Nguyen Mau and Van-Nam Huynh.
\newblock An lsh-based k-representatives clustering method for large
  categorical data.
\newblock \emph{Neurocomputing}, 463:\penalty0 29--44, 2021.
\newblock ISSN 0925-2312.
\newblock \doi{https://doi.org/10.1016/j.neucom.2021.08.050}.
\newblock URL
  \url{https://www.sciencedirect.com/science/article/pii/S0925231221012340}.

\bibitem[Han et~al.(2022)Han, Tang, and Ye]{35}
W.~Han, H.~Tang, and Y.~Ye.
\newblock Locality-sensitive hashing-based k-mer clustering for identification
  of differential microbial markers related to host phenotype.
\newblock \emph{Journal of Computational Biology}, 29\penalty0 (7):\penalty0
  738--751, Jul 2022.
\newblock \doi{10.1089/cmb.2021.0640}.
\newblock URL \url{https://doi.org/10.1089/cmb.2021.0640}.

\bibitem[Andoni and Indyk(2008)]{10.1145/1327452.1327494}
Alexandr Andoni and Piotr Indyk.
\newblock Near-optimal hashing algorithms for approximate nearest neighbor in
  high dimensions.
\newblock \emph{Commun. ACM}, 51\penalty0 (1):\penalty0 117–122, January
  2008.
\newblock ISSN 0001-0782.
\newblock \doi{10.1145/1327452.1327494}.
\newblock URL \url{https://doi.org/10.1145/1327452.1327494}.

\bibitem[Lv et~al.(2007)Lv, Josephson, Wang, Charikar, and
  Li]{10.5555/1325851.1325958}
Qin Lv, William Josephson, Zhe Wang, Moses Charikar, and Kai Li.
\newblock Multi-probe lsh: efficient indexing for high-dimensional similarity
  search.
\newblock In \emph{Proceedings of the 33rd International Conference on Very
  Large Data Bases}, VLDB '07, page 950–961. VLDB Endowment, 2007.
\newblock ISBN 9781595936493.

\bibitem[Norouzi et~al.(2014)Norouzi, Punjani, and
  Fleet]{norouzi2014fastexactsearchhamming}
Mohammad Norouzi, Ali Punjani, and David~J. Fleet.
\newblock Fast exact search in hamming space with multi-index hashing, 2014.
\newblock URL \url{https://arxiv.org/abs/1307.2982}.

\bibitem[Shen et~al.(2020)Shen, Qin, Chen, Yu, Liu, Zhu, Shen, and Shao]{tbh}
Yuming Shen, Jie Qin, Jiaxin Chen, Mengyang Yu, Li~Liu, Fan Zhu, Fumin Shen,
  and Ling Shao.
\newblock Auto-encoding twin-bottleneck hashing, 2020.
\newblock URL \url{https://arxiv.org/abs/2002.11930}.

\bibitem[Kraska et~al.(2018)Kraska, Beutel, Chi, Dean, and
  Polyzotis]{kraska2018case}
Tim Kraska, Alex Beutel, Ed~H. Chi, Jeffrey Dean, and Neoklis Polyzotis.
\newblock The case for learned index structures.
\newblock In \emph{Proceedings of the 2018 International Conference on
  Management of Data (SIGMOD '18)}, pages 489--504. ACM, 2018.
\newblock \doi{10.1145/3183713.3196909}.

\bibitem[Ding et~al.(2020)Ding, Minhas, Yu, Wang, Do, Li, Zhang, Chandramouli,
  Gehrke, Kossmann, Lomet, and Kraska]{ding2020alex}
Jialin Ding, Umar~Farooq Minhas, Jia Yu, Chi Wang, Jaeyoung Do, Yinan Li,
  Hantian Zhang, Badrish Chandramouli, Johannes Gehrke, Donald Kossmann, David
  Lomet, and Tim Kraska.
\newblock Alex: An updatable adaptive learned index.
\newblock In \emph{Proceedings of the 2020 ACM SIGMOD International Conference
  on Management of Data (SIGMOD '20)}, pages 969--984. ACM, 2020.
\newblock \doi{10.1145/3318464.3389711}.

\bibitem[Li et~al.(2020)Li, Lu, Zheng, Yang, and Pan]{li2020lisa}
Pengfei Li, Hua Lu, Qian Zheng, Long Yang, and Gang Pan.
\newblock Lisa: A learned index structure for spatial data.
\newblock In \emph{Proceedings of the 2020 ACM SIGMOD International Conference
  on Management of Data (SIGMOD '20)}, pages 2119--2133. ACM, 2020.
\newblock \doi{10.1145/3318464.3389703}.

\bibitem[Cao et~al.(2016)Cao, Long, Wang, Zhu, and
  Wen]{cao2016deepquantization}
Yue Cao, Mingsheng Long, Jianmin Wang, Han Zhu, and Qingfu Wen.
\newblock Deep quantization network for efficient image retrieval.
\newblock In \emph{Proceedings of the Thirtieth AAAI Conference on Artificial
  Intelligence (AAAI '16)}, pages 3457--3463. AAAI, 2016.
\newblock \doi{10.1609/aaai.v30i1.10313}.

\bibitem[Xiao et~al.(2022)Xiao, Liu, Han, Zhang, Lian, Gong, Chen, Yang, Sun,
  Shao, and Xie]{xiao2022distillvq}
Shitao Xiao, Zheng Liu, Weihao Han, Jianjin Zhang, Defu Lian, Yeyun Gong,
  Qi~Chen, Fan Yang, Hao Sun, Yingxia Shao, and Xing Xie.
\newblock Distill-vq: Learning retrieval oriented vector quantization by
  distilling knowledge from dense embeddings.
\newblock In \emph{Proceedings of the 45th International ACM SIGIR Conference
  on Research and Development in Information Retrieval (SIGIR '22)}, pages
  1513--1523. ACM, 2022.
\newblock \doi{10.1145/3477495.3531799}.

\bibitem[O'Neill and Dutta(2023)]{oneill2023improvedvq}
James O'Neill and Sourav Dutta.
\newblock Improved vector quantization for dense retrieval with contrastive
  distillation.
\newblock In \emph{Proceedings of the 46th International ACM SIGIR Conference
  on Research and Development in Information Retrieval (SIGIR '23)}, pages
  2072--2076. ACM, 2023.
\newblock \doi{10.1145/3539618.3592001}.

\bibitem[Zhan et~al.(2023)Zhan, Mao, Liu, Guo, Zhang, and Ma]{zhan2023repconc}
Jingtao Zhan, Jiaxin Mao, Yiqun Liu, Jiafeng Guo, Min Zhang, and Shaoping Ma.
\newblock Learning discrete representations via constrained clustering for
  effective and efficient dense retrieval (extended abstract).
\newblock In \emph{Proceedings of the Thirty-Second International Joint
  Conference on Artificial Intelligence (IJCAI '23)}, pages 6504--6508. IJCAI,
  2023.
\newblock \doi{10.24963/ijcai.2023/728}.

\bibitem[Weyand et~al.(2020)Weyand, Araujo, Cao, and Sim]{ref38}
Tobias Weyand, Andre Araujo, Bingyi Cao, and Jack Sim.
\newblock Google landmarks dataset v2 - a large-scale benchmark for
  instance-level recognition and retrieval.
\newblock In \emph{Proceedings of the IEEE/CVF Conference on Computer Vision
  and Pattern Recognition (CVPR)}, pages 2575--2584, 2020.
\newblock URL \url{https://github.com/cvdfoundation/google-landmark}.

\bibitem[Babenko and Lempitsky(2016)]{7780595}
Artem Babenko and Victor Lempitsky.
\newblock Efficient indexing of billion-scale datasets of deep descriptors.
\newblock In \emph{2016 IEEE Conference on Computer Vision and Pattern
  Recognition (CVPR)}, pages 2055--2063, 2016.
\newblock \doi{10.1109/CVPR.2016.226}.

\bibitem[Oquab et~al.(2024)Oquab, Darcet, Moutakanni, Vo, Szafraniec, Khalidov,
  Fernandez, Haziza, Massa, El-Nouby, Assran, Ballas, Galuba, Howes, Huang, Li,
  Misra, Rabbat, Sharma, Synnaeve, Xu, Jegou, Mairal, Labatut, Joulin, and
  Bojanowski]{ref43}
Maxime Oquab, Timothée Darcet, Théo Moutakanni, Huy Vo, Marc Szafraniec,
  Vasil Khalidov, Pierre Fernandez, Daniel Haziza, Francisco Massa, Alaaeldin
  El-Nouby, Mahmoud Assran, Nicolas Ballas, Wojciech Galuba, Russell Howes,
  Po-Yao Huang, Shang-Wen Li, Ishan Misra, Michael Rabbat, Vasu Sharma, Gabriel
  Synnaeve, Hu~Xu, Hervé Jegou, Julien Mairal, Patrick Labatut, Armand Joulin,
  and Piotr Bojanowski.
\newblock {DINOv2}: Learning robust visual features without supervision, 2024.
\newblock URL \url{https://arxiv.org/abs/2304.07193}.

\bibitem[Gong et~al.(2013)Gong, Lazebnik, Gordo, and Perronnin]{itq}
Yunchao Gong, Svetlana Lazebnik, Albert Gordo, and Florent Perronnin.
\newblock Iterative quantization: A procrustean approach to learning binary
  codes for large-scale image retrieval.
\newblock \emph{IEEE Transactions on Pattern Analysis and Machine
  Intelligence}, 35\penalty0 (12):\penalty0 2916--2929, 2013.
\newblock \doi{10.1109/TPAMI.2012.193}.

\bibitem[Rajput et~al.(2023)Rajput, Mehta, Singh, Keshavan, Vu, Heldt, Hong,
  Tay, Tran, Samost, Kula, Chi, and
  Sathiamoorthy]{rajput2023recommendersystemsgenerativeretrieval}
Shashank Rajput, Nikhil Mehta, Anima Singh, Raghunandan~H. Keshavan, Trung Vu,
  Lukasz Heldt, Lichan Hong, Yi~Tay, Vinh~Q. Tran, Jonah Samost, Maciej Kula,
  Ed~H. Chi, and Maheswaran Sathiamoorthy.
\newblock Recommender systems with generative retrieval, 2023.
\newblock URL \url{https://arxiv.org/abs/2305.05065}.

\bibitem[Wang et~al.(2018)Wang, Huang, Zhao, Zhang, Zhao, and
  Lee]{wang2018billionscalecommodityembeddingecommerce}
Jizhe Wang, Pipei Huang, Huan Zhao, Zhibo Zhang, Binqiang Zhao, and Dik~Lun
  Lee.
\newblock Billion-scale commodity embedding for e-commerce recommendation in
  alibaba, 2018.
\newblock URL \url{https://arxiv.org/abs/1803.02349}.

\bibitem[Xie et~al.(2023)Xie, Liu, Hou, and Huang]{ref3}
Xingrui Xie, Han Liu, Wenzhe Hou, and Hongbin Huang.
\newblock A brief survey of vector databases.
\newblock In \emph{2023 9th International Conference on Big Data and
  Information Analytics (BigDIA)}, pages 364--371, 2023.
\newblock \doi{10.1109/BigDIA60676.2023.10429609}.

\bibitem[Iscen et~al.(2017)Iscen, Furon, Gripon, Rabbat, and Jégou]{ref28}
A.~Iscen, T.~Furon, V.~Gripon, M.~Rabbat, and H.~Jégou.
\newblock Memory vectors for similarity search in high-dimensional spaces.
\newblock \emph{IEEE Transactions on Big Data}, 4\penalty0 (4), 2017.
\newblock URL \url{https://ieeexplore.ieee.org/document/7870636}.

\end{thebibliography}

\clearpage
\appendix
\def\NeuRouteCombined{}
\ifdefined\NeuRouteCombined
\else
\documentclass[11pt]{article}

\usepackage[margin=1in]{geometry}
\usepackage[T1]{fontenc}
\usepackage{graphicx}
\usepackage{booktabs}
\usepackage{multirow}
\usepackage{array}
\usepackage{tabularx}
\usepackage{float}
\usepackage{placeins}
\usepackage{amsmath}
\usepackage{amssymb}
\usepackage{adjustbox}
\usepackage{xcolor}
\usepackage{algorithm}
\usepackage{algpseudocode}
\usepackage{hyperref}
\hypersetup{
  colorlinks=true,
  linkcolor=blue,
  citecolor=blue,
  urlcolor=blue
}

\algrenewcommand\algorithmicrequire{\textbf{Input:}}
\algrenewcommand\algorithmicensure{\textbf{Output:}}

\begin{document}

\title{Appendix for NeuRoute: Logit-Guided Neural Routing for Billion-Scale Vector Search with Sub-Hour Index Construction}
\author{Xingqiao Wang \quad Zi Wang \quad Xiaowei Xu\\
\small University of Arkansas at Little Rock}
\date{}
\maketitle
\appendix
\fi

\numberwithin{figure}{section}
\numberwithin{table}{section}
\numberwithin{algorithm}{section}

\section{NeuRoute model configurations}
\label{app:encoder_config}

Table~\ref{tab:NeuRoute-arch} summarizes the encoder instantiations used in our evaluation.
Across all settings, NeuRoute uses a lightweight MLP encoder that progressively compresses the input embedding from $E_{\mathrm{dim}}$ to a compact latent logit vector of dimension $L_{\mathrm{dim}}$.
Each hidden layer applies a linear layer followed by BatchNorm and ReLU, while the final encoder layer remains linear to preserve the latent logits used for binarization and routing.
We also report the dataset-specific $\tau_{\mathrm{emb}}$ used for top-quantile pair selection; for a given dataset family (e.g., BigANN), we use the same $\tau_{\mathrm{emb}}$ across scales.

\begin{table}[H]
  \caption{\textbf{NeuRoute model design.} Hidden layers use Linear $\rightarrow$ BatchNorm $\rightarrow$ ReLU; the final encoder layer is linear (no BatchNorm or activation).}
  \label{tab:NeuRoute-arch}
  \centering
  \small
  \setlength{\tabcolsep}{4pt}
  \renewcommand{\arraystretch}{1.05}
  \begin{adjustbox}{max width=\linewidth}
  \begin{tabular}{@{}lccccc@{}}
    \toprule
    \textbf{Dataset} & \textbf{$E_{\mathrm{dim}}$} & \textbf{$L_{\mathrm{dim}}$} & \textbf{$\gamma$} & \textbf{$\tau_{\mathrm{emb}}$} & \textbf{Encoder widths} \\
    \midrule
    \texttt{BigANN-100M}  & 128  & 20 & 0.6 & 1.20 & $128, 96, 64, 20$ \\
    \texttt{BigANN-1B}    & 128  & 22 & 0.6 & 1.20 & $128, 96, 64, 22$ \\
    \texttt{Deep1B-100M}  & 96   & 20 & 0.6 & 0.85 & $96, 64, 32, 20$ \\
    \texttt{Deep1B-1B}    & 96   & 22 & 0.6 & 0.85 & $96, 64, 32, 22$ \\
    \texttt{GLDv2}        & 1536 & 16 & 0.6 & 0.90 & $1536, 512, 256, 128, 64, 16$ \\
    \bottomrule
  \end{tabular}
  \end{adjustbox}
\end{table}

\FloatBarrier
\section{Training stability and convergence}
\label{app:train_val_loss}

\begin{figure}[t]
  \centering
  \includegraphics[width=\linewidth]{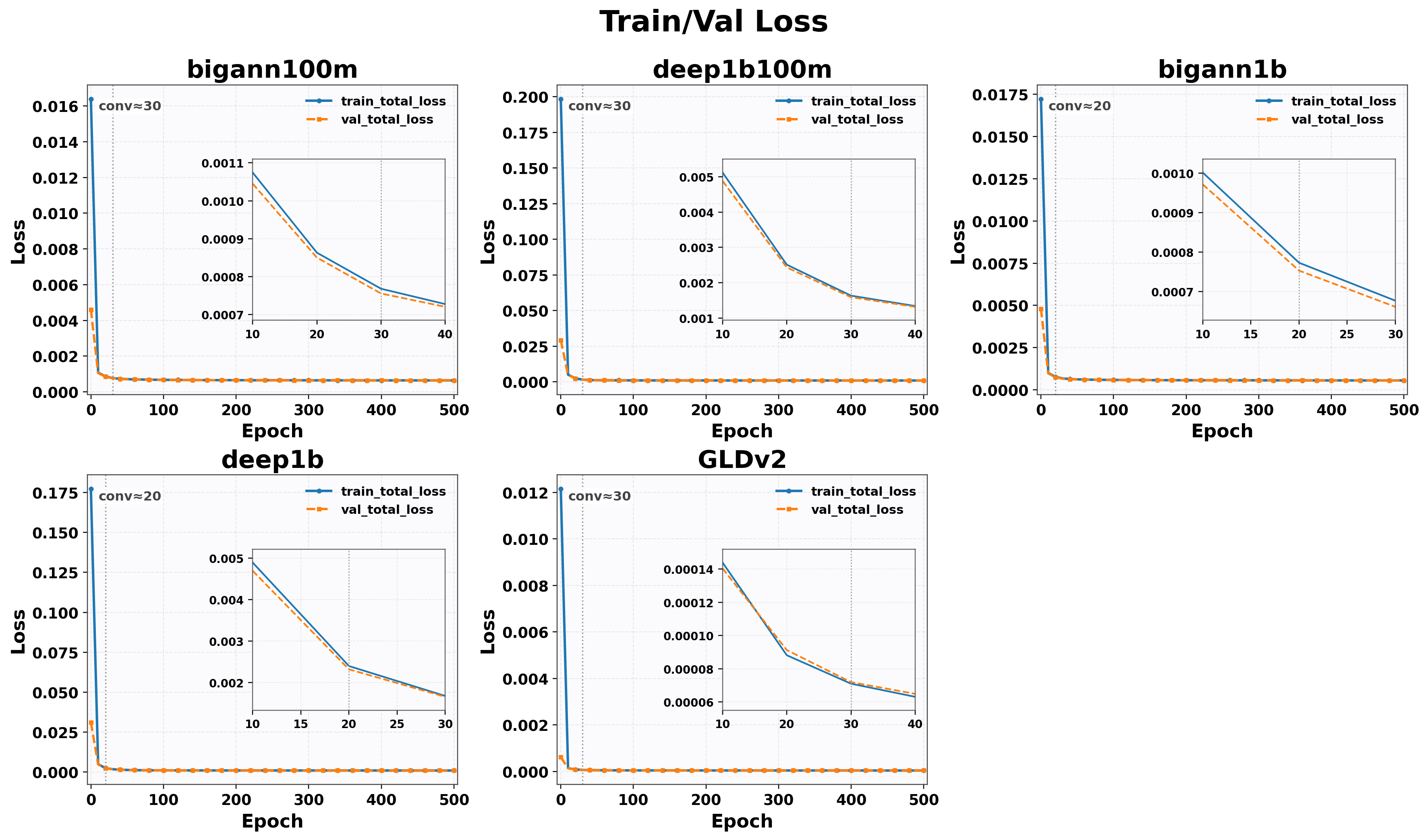}
  \caption{NeuRoute training/validation loss curves across datasets. Each panel shows the total loss over epochs with an inset zooming into the late-epoch regime to highlight convergence.}
  \label{fig:train_val_loss}
\end{figure}

Figure~\ref{fig:train_val_loss} reports the training and validation loss curves for NeuRoute across all evaluated datasets under the default training protocol (2M training subset, batch size 4096, 500 epochs).
Across datasets, the loss decreases rapidly in early epochs and then stabilizes, indicating consistent convergence behavior under a fixed schedule; we use this fixed schedule for all experiments for simplicity and reproducibility.

\FloatBarrier
\section{Empirical evidence for logit-guided bucket routing}
\label{app:logit_routing_evidence}

\begin{figure}[t]
  \centering
  \includegraphics[width=\columnwidth]{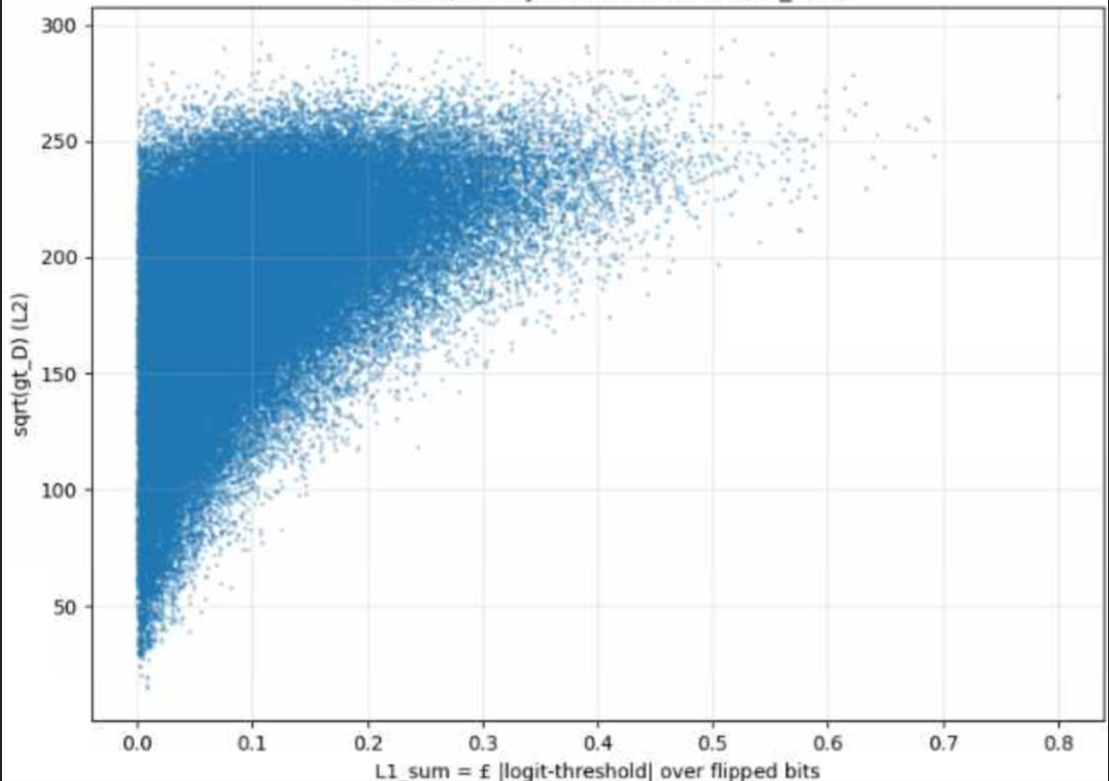}
  \caption{KNN embedding distance versus the proposed bucket score (\texttt{L1\_sum}). Each point corresponds to a probed bucket; lower scores tend to be associated with closer GT neighbors.}
  \label{fig:l1sum-vs-gt}
\end{figure}

\begin{figure}[t]
  \centering
  \includegraphics[width=\columnwidth]{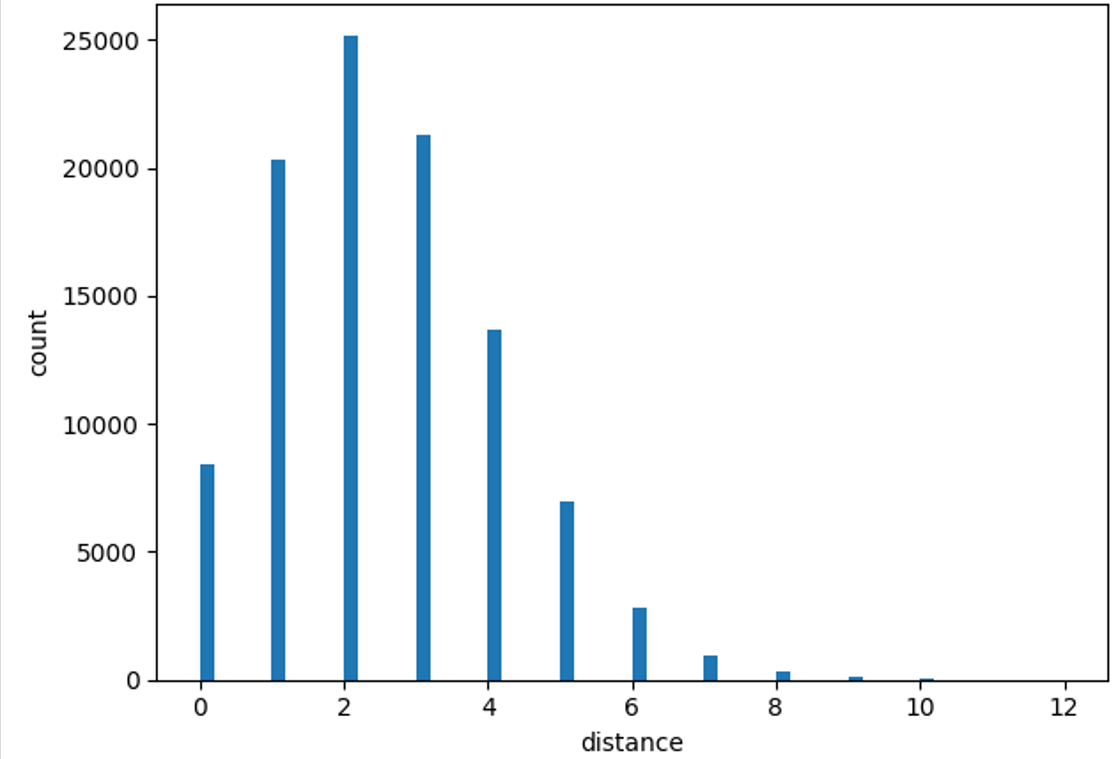}
  \caption{Distribution of query--candidate Hamming distances observed during probing. Useful candidates concentrate at small Hamming distances, motivating prioritized multi-bucket enumeration under a fixed budget.}
  \label{fig:q-c-hamming}
\end{figure}

\begin{figure}[t]
  \centering
  \includegraphics[width=\columnwidth]{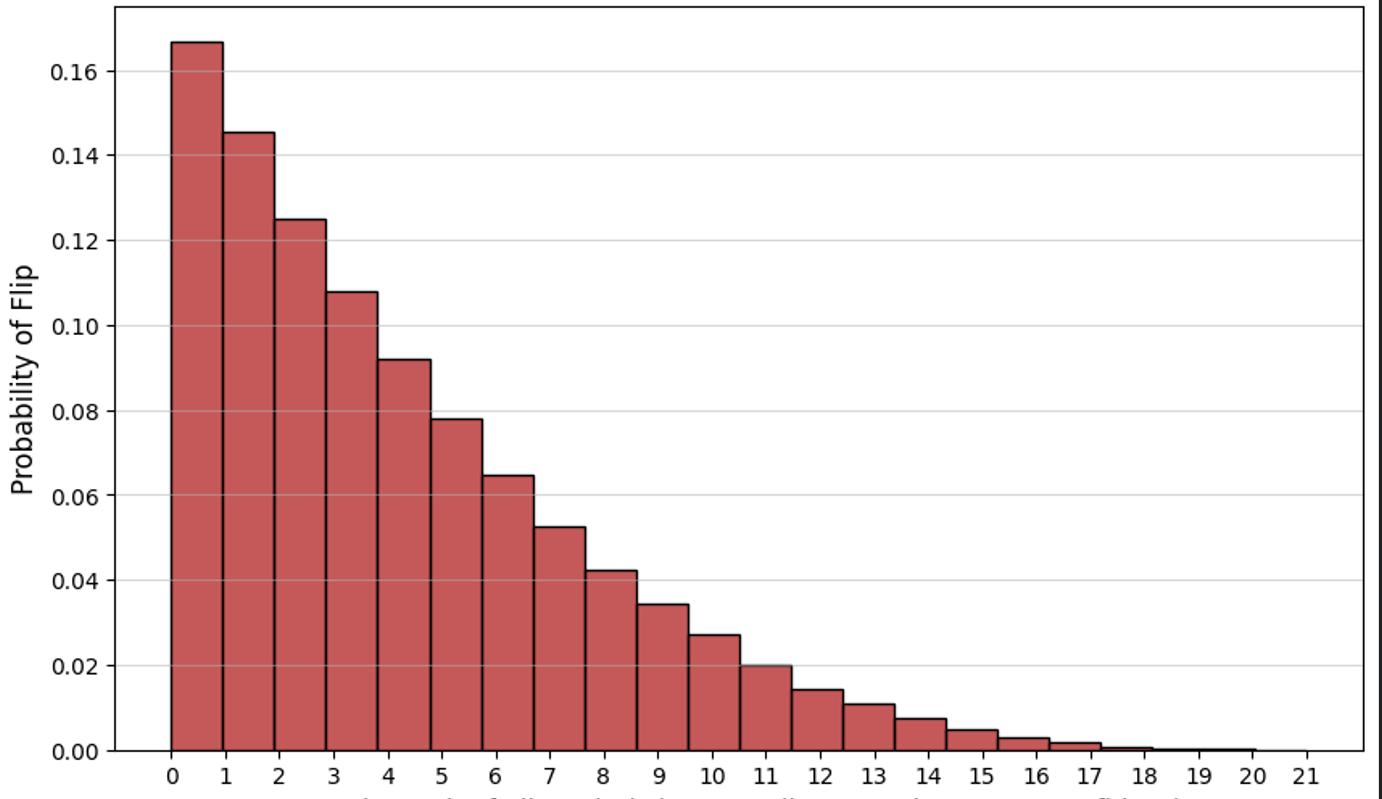}
  \caption{Probability that a KNN flips a bit as a function of the bit's margin rank (rank 0 = smallest margin / least confident). GT flips concentrate on low-margin bits, validating margin-ranked bit flipping for bucket routing.}
  \label{fig:gt-flip-rank}
\end{figure}

We focus on \emph{bucket routing/refinement} rather than improving hash codes.
Given fixed codes and a fixed hash-table layout, we exploit the pre-binarization logit margins to prioritize which buckets to probe under a limited budget.
Figures~\ref{fig:l1sum-vs-gt}, \ref{fig:q-c-hamming}, and \ref{fig:gt-flip-rank} support this design choice from three angles:
(i) our bucket score correlates with KNN distance, (ii) useful candidates concentrate at small Hamming distances, and (iii) KNN bit flips are biased toward low-margin (uncertain) bits.

\FloatBarrier
\section{NeuRoute retrieval and build breakdown at \texorpdfstring{$\sim$}{approximately }0.90 Recall@10}
\label{app:neuroute_breakdown_90}

\begin{figure}[t]
  \centering
  \includegraphics[width=0.80\textwidth]{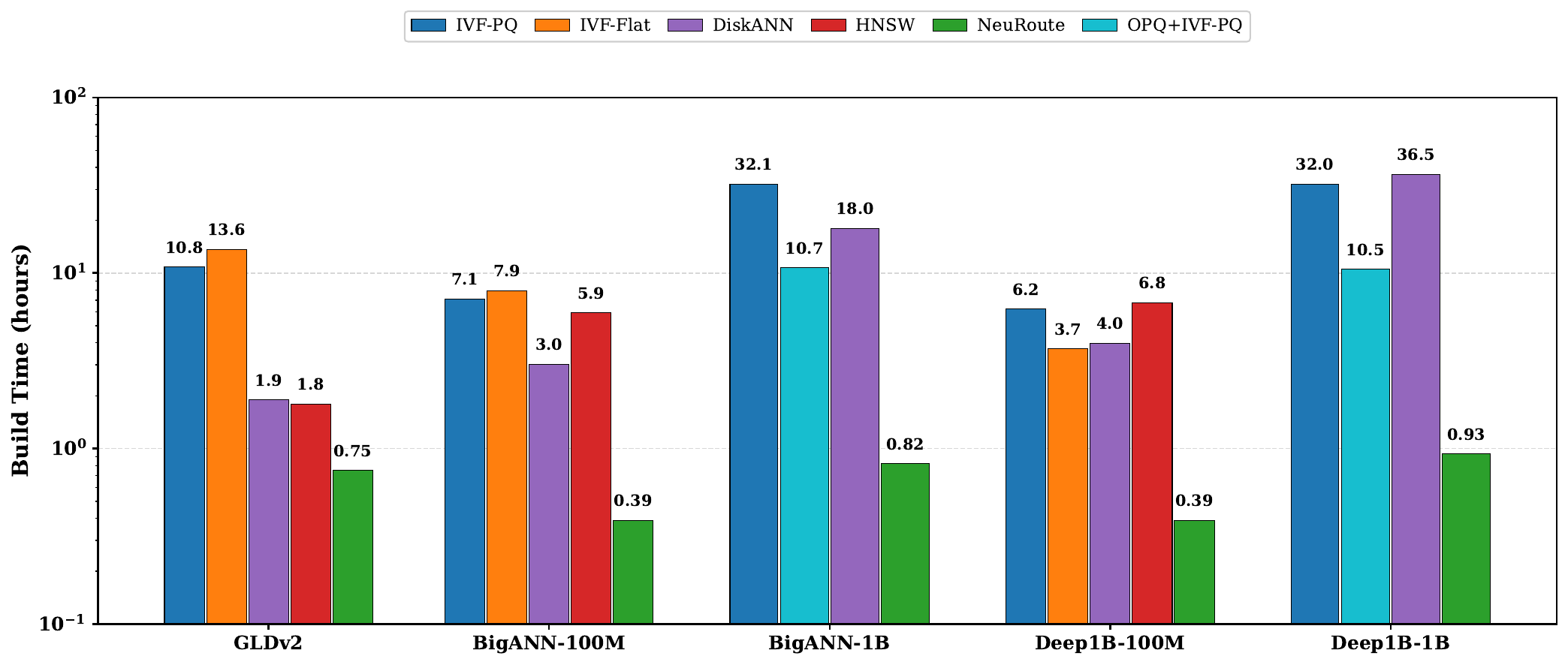}
  \caption{End-to-end index build time (hours, log-scale) for IVF--PQ, IVF--Flat, DiskANN, HNSW, OPQ + IVF--PQ, and NeuRoute.}
  \label{fig:buildtime}
\end{figure}

This section reports NeuRoute's query-time work decomposition at operating points with Recall@10 closest to 0.90 (Table~\ref{tab:neuroute_retrieval_breakdown_90}), including bucket-enumeration time, refinement time, and the resulting candidate volume per query.
Together with the end-to-end build-time comparison in Fig.~\ref{fig:buildtime}, these results highlight NeuRoute's intended design point: budget-controlled retrieval with a lightweight time-to-index pipeline under a single-node setup.

\begin{table}[H]
\centering
\setlength{\tabcolsep}{3pt}
\renewcommand{\arraystretch}{1.05}
\caption{NeuRoute retrieval breakdown at operating points with Recall@10 closest to 0.90 on each dataset. \texttt{enum\_ms}, \texttt{refine\_ms}, \texttt{total\_ms}, \texttt{total\_qps}, and \texttt{recall} are reported from query-level summary logs. \texttt{routing\_ms}$=\texttt{total\_ms}-\texttt{enum\_ms}-\texttt{refine\_ms}$. \texttt{Cand./q} is the average number of refined candidates per query, computed as \texttt{refined\_vecs}/$Q$.}
\label{tab:neuroute_retrieval_breakdown_90}
\begin{adjustbox}{max width=\textwidth}
\begin{tabular}{l r r r r r r r r}
\toprule
Dataset & \texttt{enum\_ms} & \texttt{refine\_ms} & \texttt{routing\_ms} & \texttt{total\_ms} &
\texttt{total\_qps} & Cand./q & \texttt{refined\_vecs} & \texttt{recall} \\
\midrule
\texttt{BigANN-1B}   & 989.62  & 2832.63  & 31.51  & 3853.76  & 2594.87 & 193{,}591 & \ensuremath{1.94\times 10^{9}} & 0.89965 \\
\texttt{Deep1B-1B}   & 1017.18 & 13022.58 & 45.46  & 14085.22 & 709.96  & 418{,}079 & \ensuremath{4.18\times 10^{9}} & 0.90000 \\
\texttt{BigANN-100M} & 268.98  & 1743.22  & 33.22  & 2045.42  & 4888.96 & 142{,}093 & \ensuremath{1.42\times 10^{9}} & 0.90059 \\
\texttt{Deep1B-100M} & 294.48  & 7110.12  & 38.04  & 7442.64  & 1343.61 & 227{,}449 & \ensuremath{2.27\times 10^{9}} & 0.90006 \\
\texttt{GLDv2}       & 34.12   & 12718.24 & 220.81 & 12973.17 & 770.82  & 27{,}041  & \ensuremath{2.70\times 10^{8}} & 0.90066 \\
\bottomrule
\end{tabular}
\end{adjustbox}
\vspace{-2mm}
\end{table}

\FloatBarrier
\section{GPU-only reference (CAGRA)}
\label{app:cagra_reference}

We evaluated the GPU graph baseline CAGRA as a throughput reference.
In our setup, CAGRA is limited by GPU memory and can be constructed only up to 80M vectors on an A100 40GB, so we do not include it in the main Recall@10--QPS plots that focus on 100M/1B-scale comparisons.

\begin{figure}[H]
  \centering
  \includegraphics[width=\linewidth]{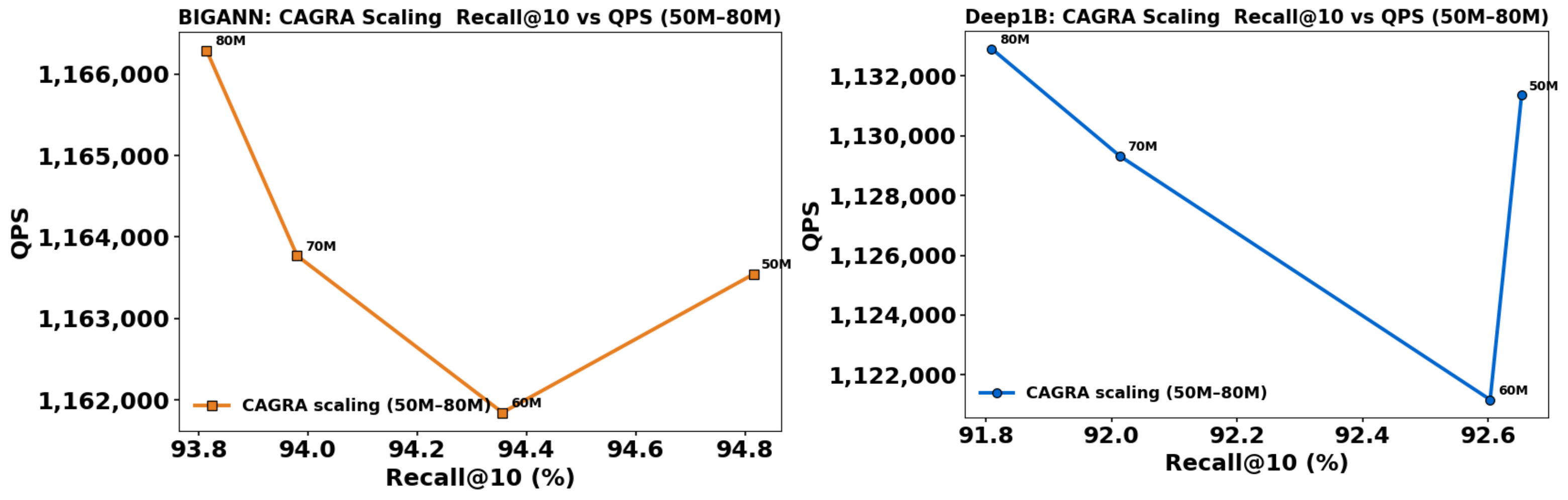}
  \caption{CAGRA GPU-only scaling reference at 50M--80M on BigANN and Deep1B. Points are labeled by dataset size (50M/60M/70M/80M).}
  \label{fig:cagra_scaling_50_80m}
\end{figure}

At the 80M scale, CAGRA achieves very high throughput while maintaining strong recall: on BigANN-80M it reaches Recall@10 $=0.938$ at $\sim 1.17\times 10^{6}$ QPS, and on Deep1B-80M it reaches Recall@10 $=0.918$ at $\sim 1.13\times 10^{6}$ QPS.
These results highlight the effectiveness of GPU graph search under sufficient device memory, but its scaling is constrained in our environment.

\FloatBarrier
\ifdefined\NeuRouteCombined
\else
\end{document}
\fi

\end{document}